\documentclass[amssymb,amsmath,twocolumn,showpacs,aps,pra,floatfix]{revtex4-1}

\usepackage{color,graphicx,bm}
\usepackage{overpic}
\usepackage{upgreek}

\def\k{{\bk}}
\def\q{{\bq}}
\def\b #1{{\bf #1}}
\def\t #1{{\text{#1}}}

\def\beq{\begin{equation}}
\def\eeq{\end{equation}}
\def\nn{\nonumber}

\def\Om{\Omega}

\def\lam{\lambda}
\def\th{\theta}
\def\vth{\vartheta}
\def\eps{\epsilon}

\def\s{\sigma}

\def\vp{\varphi}
\def\p{\phi}

\def\rd{{\rm{d}}}

\def\Zz{\mathbb{Z}}
\def\Rr{\mathbb{R}}

\def\pa{\partial}

\def\sign{{\rm sign}}

\def\bv{\bm{v}}

\def\bk{\bm{k}}
\def\bq{\bm{q}}
\def\bh{\bm{h}}
\def\bsi{\bm{\s}}
\def\bu{\bm{u}}
\def\bOm{\bm{\Omega}}
\def\be{\bm{e}}
\def\bnull{\bm{0}}
\def\bs{\bm{s}}
\def\bS{\bm{S}}

\def\bH{\bm{H}}

\def\hJ{\hat{J}}

\def\Re{\textrm{Re}}
\def\Im{\textrm{Im}}
 
\begin{document}

\title{Momentum-Space Spin Texture in a Topological Superconductor\\[0mm]
}

\author{
Florian Loder$^{\,1,2}$, Arno P. Kampf$^{\,2}$, Thilo Kopp$^{\,1}$, and Daniel Braak$^{\,1}$
\vspace{0,3cm}}

\affiliation{Center for Electronic Correlations and Magnetism, $^1$Experimental Physics VI, 
$^2$Theoretical Physics III\\ 
Institute of Physics, University of Augsburg, 86135 Augsburg, Germany}

\date{January 3, 2017}

\begin{abstract}
A conventional superconductor with spin-orbit coupling turns into
a topological superconductor beyond a critical strength of the
Zeeman energy. The spin-expectation values $\bS(\bk)$ in momentum space
trace this transition via a characteristic change in the topological
character of the spin texture within the Brillouin zone. At the
transition the skyrmion counting number switches from 0 to 1/2
identifying the topological superconductor via its meron-like spin
texture. The change in the skyrmion counting number is crucially
controlled by singular points of the map $\bS(\bk)/|\bS(\bk)|$ from the
Brillouin zone, i.e.\ a torus, to the unit sphere. The complexity
of this spin-map is discussed at zero temperature as well as for
the extension to finite temperatures.
\end{abstract}


\maketitle

\section{Introduction}
\label{sec:intro}

The classification of electronic states by integer numbers that are not quantum 
numbers but derive from the topology of the entire electronic system has become a  
significant concept in the past two decades~\cite{thouless:98,bernevig:13,hasan:10,qi:11}. 
Whereas the quantum 
numbers may depend on the details of the realization of a complex system, the topological 
character is considered to be robust. The topology is a global property of the entire set 
of the system's quantum states. In many electronic systems, the robustness results from 
an energy gap that separates the occupied from the unoccupied states. Closing and 
reopening this gap by the variation of control parameters may allow for a transition into a 
topologically different state.  Solids 
that are insulators or superconductors can be classified accordingly. 
A prominent example for a topological insulator is the quantum Hall state, where 
the number of occupied Landau levels defines its topological character~\cite{ezawa:13}. 
Thouless, Kohmoto, Nightingale, and den Nijs showed that the corresponding topological number is obtained 
as an integral of the Berry curvature over the Brillouin zone (BZ) 
~\cite{thouless:82}. 
It is typically referred to as the TKNN number $C_\t{TKNN}$. In mathematically precise terms
the curvature is the first Chern class and $C_\t{TKNN}$ the first Chern number
of the principal $U(1)$ bundle of wave functions over the torus $T^2$ of momentum
eigenstates in the BZ~\cite{kohmoto:82}.

Also a nodeless superconductor can be characterized by a topological invariant. 
Topologically similar to the quantum Hall state is a two-dimensional (2D) $s$-wave superconductor 
with a  Zeeman coupling to a sufficiently strong magnetic field in the presence 
of a Rashba spin-orbit coupling (SOC)~\cite{sau:10,alicea:10,sato2:09,sato:10,tewari:11,gong:12}. 
The same physics applies to topological superfluidity in polarized ultracold atomic
Fermi gases with SOC~\cite{zhou:11,seo:12,hu:11,qu:13,zhang:13,chen:13}.
The topological $s$-wave superconductor is sometimes referred 
to as an effective $p$-wave superconductor, because its pairing amplitude
in the two spin-orbit split bands acquires the antisymmetric wavevector-dependence of 
the Rashba SOC~\cite{alicea:10}. Nevertheless, its topological properties are different from those of 
a true triplet $p$-wave superconductor. In particular, whereas the $s$-wave superconductor 
remains topologically trivial in the absence of a magnetic field, the $p$-wave superconductor itself
has a non-trivial topology, which is not captured by $C_\t{TKNN}$, but rather by a $\mathbb Z_2$ 
topological quantum number $C_{\mathbb Z_2}$ that applies to  time-reversal invariant
systems~\cite{kane1:05,kane2:05,moore:07,fu:07,roy:08,kitaev:09,qi:09,sato1:09}.

While the non-trivial topology of the $p$-wave state is intrinsic to the spin structure 
defined by the superconducting (SC) order parameter (OP), a topologically non-trivial
character of a  conventional $s$-wave 
superconductor requires that the normal conducting state already exhibits 
a specific spin structure in momentum space. This spin texture is 
imposed by the SOC; its relation to the topological character 
is, however, by no means obvious. To analyze the emergence of 
the topological SC state 
the Chern and the skyrmion numbers was previously introduced also in Ref.~\cite{annica:14}.

However, although the spin texture
in momentum space is reminiscent of a skyrmion,
its pattern might not completely cover the full solid angle of all possible spin orientations.
If the texture covers half of the
full solid angle, it may be interpreted as a kind of ``half skyrmion'',
which for spatial spin patterns has been called a meron~\cite{volovikbook}. 
But is the spin structure of such a meron  
characterized by a well-defined topological invariant? The resolution of this fundamental issue is provided by
our analysis. We find that the mapping of the BZ (torus) to the hemisphere of the normalized 
spin expectation values becomes singular at a finite number of points in the BZ; the analysis of these
singular points  allows us to prove the topological character of the meron spin structure.

These findings naturally pose the question, if and how the topological character of
the $s$-wave superconductor is also reflected in its finite temperature behavior. 
The Berry curvature and the respective Chern number are, by construction, meant to reveal
intrinsic features of the system's groundstate and therefore they are unsuited to address finite temperatures.
Moreover, a topological invariant, such as the skyrmion number, cannot reflect a smooth evolution 
with temperature as the topological invariant is constrained to integer numbers. We investigate instead 
if the Hall conductance or deliberately selected measures of the spin texture provide tools to
continuously follow the evolution from the topological ground state to the canonical ensemble at
finite temperatures. Specifically, we identify a peculiar spin product which serves this purpose.
At zero temperature it is equivalent to the Berry curvature but, since it is based on expectation
values of spin operators, this spin product is straightforwardly extended to finite temperatures.

For a superconductor with SOC the orientation of the external magnetic field matters, and we therefore
also examine the momentum-space spin texture upon field rotation. As a sufficiently strong in-plane component 
of the field induces finite-momentum pairing, this analysis is intricate. Yet, we show that the crucial vortex
structures, which are attached to the singular points in the spin map from the BZ to the unit sphere, are 
preserved also for an in-plane magnetic field component. The meron character of the spin texture
in the 2D topological superconductor is therefore maintained, when the magnetic field is rotated from
an out-of-plane to an in-plane orientation.

\section{Topological concepts}
\label{sec:concepts}
Before analyzing the invariants for a topological superconductor  we first recollect some
of the basic concepts for topological insulators, which the subsequent sections build on.
Two-band insulators are straightforward generalizations of two-level systems which
can be written in terms of pseudospins represented by Pauli matrices in the Hamiltonian. 
The real-valued Bloch vector $\bh$ controls the rotation in the pseudospin space.

The two-band insulator is a paradigmatic electronic system to introduce topological quantum numbers.
In this section we relate seemingly different topological invariants of the electronic state. 
Specifically we will address how the Chern number is related to the Brouwer degree of a map from a torus $T^2$ to 
a unit sphere $S^2$ and, moreover, 
how this is related to the spin texture in momentum space and the associated skyrmion number.
Eventually, the Chern number may be identified as the TKNN number which can be derived from a Kubo formula 
for the Hall conductance.
These different topological aspects have been discussed in several textbooks
(see, for example, the books by Thouless~\cite{thouless:98}, Bernevig~\cite{bernevig:13}, Volovik~\cite{volovikbook}, Jost~\cite{jost}, 
and references therein).
In this section we shall connect the aspects raised above. It is Stokes' theorem for multiply connected surfaces
which is pivotal in deriving these relations, especially the connection between the Chern number and the Brouwer degree of the map generated by the normalized Bloch vector~\cite{bott-tu}. 

In two space dimensions we consider the two-band lattice Hamiltonian, diagonal in momentum space,
  \beq\label{2bandH}
  H=\sum_{\bk} [\eps_{\bk}\s^0 + \bh({\bk})\cdot\bsi]_{ss'}c^\dagger_{\bk s}c^{\phantom\dagger}_{\bk s'}
  \eeq
The pseudospin indices $s,s'$  for the electron creation and annihilation operators refer 
to a discrete degree of freedom. Later on we will specify their actual
nature (spin and/or orbital degrees of freedom); $\bsi$ denotes the vector of Pauli matrices.
The Bloch vector $\bh({\bk})$ is assumed to be a smooth, periodic, nowhere vanishing function of 
the wavevector $\bk=(k_1,k_2)^\top$, where $\bk$ varies through the two-dimensional Brillouin zone 
$-\pi\le k_j\le\pi$. 

It describes therefore a smooth map from the torus $T^2$ (a compact manifold without boundary) into $\Rr^3$. The normalized Bloch 
vector $\bh({\bk})/h({\bk})$ with $h({\bk})=|\bh({\bk})|$ is a map from $T^2$ into the unit sphere $S^2$.
The eigenenergies of the two bands are $\xi_{\pm,{\bk}}=\eps_{\bk}\pm |\bh({\bk})|$ and
the normalized single-particle eigenstates of $H$ are
  \beq
  \bu_{\pm}(\bk)=\frac{1}{\sqrt{2h(h\pm h_z)}}(h_z\pm h,h_x+ih_y)^\top.
  \label{spinor}
  \eeq
Here, $\pm$ denotes the band index for the diagonalized Hamiltonian.
The $\bk$ dependence of $h$ and $h_x$, $h_y$, $h_z$
on the right hand side has been suppressed. 
In the following we consider the insulating case of a completely 
filled lower band $\bu_-(\bk)=(u_1(\bk),u_2(\bk))^\top$ and an empty upper band $\bu_{+}(\bk)$.

\subsection{Berry connection and Berry curvature}
The Berry connection of the lower band is defined as the 1-form
\beq
  \bv=i\bu^\dagger_- \,(\pa_{k_1} \bu_-)\,\rd k_1 +
          i\bu^\dagger_- \,(\pa_{k_2} \bu_-)\,\rd k_2.
\label{connection}
\eeq
The 1-form $\bv$ is real-valued due to the normalization of $\bu_-$
and $\bu_-^\dagger=(u_1^{*},u_2^{*})^\top$.
The Berry phase picked up along a path $\gamma$ in the BZ is 
  \beq
  \p(\gamma)=\int_{\gamma}\bv.
  \eeq
Specifically for the lower-band eigenstates in Eq.~(\ref{spinor}), with $\bv=v_1\rd k_1+v_2\rd k_2$ and $\pa_{k_j}=\pa_{j}$,
we obtain
  \beq
  v_j=R^{-2}(h_{y}\pa_{j}h_{x}-h_{x}\pa_jh_y),\quad R=\sqrt{2h(h-h_z)}.
  \label{conn}
  \eeq
$\bv$ is a scalar-valued 1-form, that is, each component $v_j$ transforms as a scalar $U(1)$ gauge field because the wavefunctions $\bu_-(\bk)$ are sections of a complex line bundle over the torus. The 
associated curvature is therefore the exterior derivative~\cite{jost},
  \beq
  \label{BerryC}
  \bOm=\rd\bv=(\pa_1v_2-\pa_2v_1)\rd k_1\wedge\rd k_2.
  \eeq
With
$\langle\bu_-|\nabla_{\bk}|\bu_-\rangle = (\langle\bu_-|\pa_1|\bu_-\rangle,\langle\bu_-|\pa_2|\bu_-\rangle,0)^\top$, 
we may write the value $\Omega(\bk)$ of $\bOm$ at $\bk$ 
in the familiar vectorial notation~\cite{xiao},
\beq
\Omega(\bk)=i[\nabla_{\bk}\times\langle\bu_-|\nabla_{\bk}|\bu_-\rangle]_z.
\label{berryvec}
\eeq
The 2-form $\bOm/(2\pi)$ is the first (and top) Chern class, an element of the second cohomology group $H^2(T^2)$ of the torus~\cite{footnote2}. 
Its integral over the BZ yields the first Chern number of the $U(1)$-bundle represented by $\bv$. 
If $\bv$ were a smooth function throughout the $BZ$, this integral 
would necessarily vanish. But as seen from Eq.~\eqref{conn}, $\bv$ has isolated singular points, 
if $\bh=h_z\be_z$ with $h_z>0$, that is, whenever the point $\bk$ is mapped by $\bh/h$ 
to the north pole of $S^2$ where $R(\bk)=0$. In this case one has to apply Stokes' theorem for 
multiply connected surfaces~\cite{bott-tu},
  \beq
  \int_{BZ}\bOm=\oint_{\pa BZ}\bv-\sum_l\oint_{c_l}\bv  = -\sum_l\oint_{c_l}\bv,
  \label{intBZ}
  \eeq
The paths $c_l$ denote infinitesimally small circles around the singular points $\bk_l$ with $R(\bk_l)=0$ 
which constitute punctures in the base manifold (that is, the BZ-torus).

We show in Appendix~\ref{app1} that Eq.~\eqref{intBZ}  measures a topological property of the map $\bh(\bk)$,
 \beq
  \int_{BZ}\bOm=2\pi\sum_l\sign(J(\bk_l))=2\pi C_{\rm Brouwer}.
  \label{brouwer1}
  \eeq
Here, $J(\bk_l)$ denotes the Jakobian of $\bh(\bk)$ at the singular point $\bk_l$.  The integral over the Berry curvature $\bOm$ therefore equals $2\pi$ times the Brouwer 
degree of the map $\bh(\bk)/h(\bk)$ from $T^2$ to $S^2$.  

The mapping degree $C_{\rm Brouwer}$
is a homotopy invariant of continuous maps between orientable manifolds~\cite{milnor}. For the simple case of maps from $S^1$ to $S^1$, it is just the winding number. In higher dimensions, and if base and image manifold are topologically the same, $C_{\rm Brouwer}$ can be visualized by the number of ``wrappings'' associated with the map. Here,
however, the situation is different because the maps go from $T^2$ to $S^2$ and even turn out to be singular, i.e. they are
not continuous everywhere. Nevertheless, as we show in Sec.~\ref{sec:textures}, a simple visualization is possible.  

\subsection{Skyrmion number}
\begin{figure}[t!]
\centering
\vspace{3mm}
\begin{overpic}
[width=0.98\columnwidth]{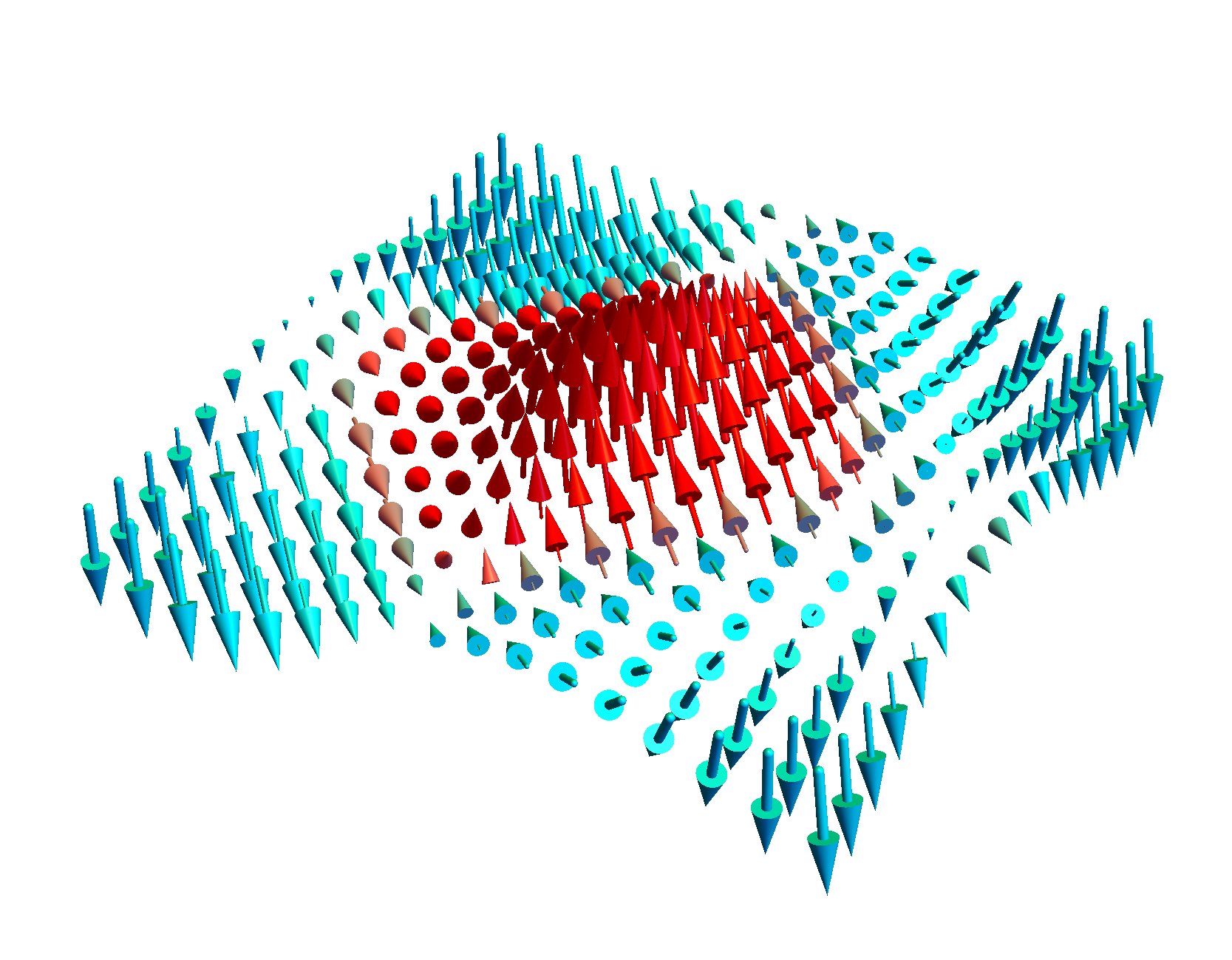}
\end{overpic}
\vspace{-0mm}
\caption{(Color online) Skyrmion texture of the Bloch vector $\bh({\bk})$ of a two-dimensional two-band 
topological insulator. Specifically, $\bh(\bk)=(\alpha \sin k_y, -\alpha \sin k_x, M+B\,[2-\cos k_x -\cos k_y])$
is plotted in the Brillouin zone of a square lattice for $\alpha=B=1$ and $M=-2$~\cite{qi:06,bernevig:13}.}
\label{Blochvector}
\end{figure}

Equation~\eqref{brouwer1} can be alternatively derived by directly evaluating $\bOm$. 
With 
\beq
\pa_1v_2-\pa_2v_1=\Om,
\label{BerryOm}
\eeq 
 and the shorthand notation
\beq
  J_{ab}=(\pa_1h_a)(\pa_2h_b)-(\pa_1h_b)(\pa_2h_a)
\label{nota}
\eeq
for $a,b\in\{x,y,z\}$, we find (see Appendix~\ref{app1})
\beq
\Om = \frac{1}{2}\frac{1}{h^3}[h_xJ_{yz}+h_yJ_{zx}+h_zJ_{xy}],
\label{Om1}
\eeq
$\Om(\bk)$ is therefore identified as a smooth function of $\bk$ if $\bh$ is smooth and vanishes nowhere, i.e. the divergences of $\bv$ have been lifted in $\Om$. 
We rewrite the Berry curvature also in the standard vectorial notation
\beq
\Om(\bk)=\frac{1}{2}\frac{\bh}{h^3}\cdot[\pa_1\bh\times\pa_2\bh].
\label{Om2}
\eeq
Two times the Berry curvature is thus
\beq\label{skyrmdensity}
\Om_{\rm h}(\bk)=\hat{\bh}\cdot[\pa_1\hat{\bh}\times\pa_2\hat{\bh}],
\eeq
where we have introduced the normalized $\hat{\bh}=\bh/h$ and used Eq.~(\ref{omega_norm}) (see Appendix~\ref{app1}).
This $\Om_{\rm h}$ is associated to another, seemingly different topological invariant, the skyrmion number. 
In fact, for a two-band topological insulator described by the Hamiltonian of Eq.~(\ref{2bandH}) with a special
choice of ${\bh}(\bk)$, the Bloch-vector field over the BZ represents a skyrmion (see Fig.~\ref{Blochvector}), and
the skyrmion number is defined as~\cite{bernevig:13,volovikbook}
\beq\label{NS}
N_{\rm S}=\frac{1}{4\pi}\int_{BZ}\Om_{\rm h}(\bk)\rd^2 k.
\eeq
$N_{\rm S}$ is necessarily an integer and equals $C_{\rm Brouwer}$, as shown in Appendix~\ref{app1}.

\subsection{Spin texture}
The $\bk$-dependent groundstate expectation value of the pseudospin
\beq
\bS(\bk)=\frac{1}{2}\langle c^\dagger_{\bk s}\bsi_{ss'}c_{\bk s'}\rangle
\eeq
reads---obtained with the normalized eigenspinors $\bu_-(\bk)$ in Eq.~\eqref{spinor}:
\begin{align} \label{spintexture}
  S_x(\bk)=&\Re[u_1^\ast(\bk)u_2(\bk)]=\frac{h_x}{2h},\nn\\
  S_y(\bk)=&\Im[u_1^\ast(\bk)u_2(\bk)]=\frac{h_y}{2h},\\
  S_z(\bk)=&\frac{1}{2}[|u_1(\bk)|^2 + |u_2(\bk)|^2]=\frac{h_z}{2h}.\nn
\end{align}
Therefore, $\bS(\bk)=\bh(\bk)/2h(\bk)$, and the topological invariant is equivalently computed 
using the spin expectation value. We write with the normalized $\hat{\bS}=\bS/|\bS|$
\beq
\label{spinkyrmionnumber}
N_{\rm S}=\frac{1}{4\pi}\int_{BZ}\hat{\bS}\cdot[\pa_1\hat{\bS}\times\pa_2\hat{\bS}]\,\rd^2 k
\eeq
with the $\bk$-dependence of $\hat{\bS}(\bk)$ suppressed.
Obviously, the Berry curvature can be expressed in terms of the expectation
value of the spin and, correspondingly, we introduce
\beq
\label{spinberrycurvature}
\Omega_{\rm S}(\bk)= \hat{\bS}\cdot[\pa_1\hat{\bS}\times\pa_2\hat{\bS}].
\eeq
As $\Omega_{\rm S}(\bk) =\Omega_h(\bk)$, 
the spin texture within the BZ is therefore, up to normalization, equivalent to the Bloch vector field
displayed in Fig.~\ref{Blochvector}. 

Here we have already denoted $\bS(\bk)$ as ``spin'' rather than pseudospin.
Indeed, it is irrelevant whether this discrete degree of freedom is a pseudospin or the true electron spin,  
as long as we do not introduce additional interaction terms.  In the spin language, the Bloch vector 
$\bh(\bk)$ parametrizes the spin-orbit coupling and the Zeeman coupling to an external magnetic field.
In the subsequent sections we will address a model superconductor that has spin degrees of freedom
included naturally.

\subsection{Kubo formula}
\label{sec:Kubo}

Within linear response theory, the Hall conductance $\sigma_{xy}=(e^2/h)\,C$  
is determined by the Kubo formula,
\begin{multline}
C=\frac{i}{2\pi}\int_{BZ}\!\rd^2 k\,\sum_{n\neq m}\frac{f(E_{n{\bk}})-f(E_{m{\bk}})}{(E_{n{\bk}}-E_{m{\bk}})^2}
\\\times\langle n,{\bk}|\hat J_x|m,{\bk}\rangle\langle m,{\bk}|\hat J_y|n,{\bk}\rangle.
\label{Kubo}
\end{multline}
The eigenenergies are labeled by $n$ (or $m$) and momentum $\bk$;  
$\hat J_x$, $\hat J_y$ are the components of the
paramagnetic current operator (in units of $e/h$), and $f(E)$ denotes the Fermi function.
In the zero temperature limit, $C=C_\t{TKNN}$, the first Chern number.  This is verified by explicitly
evaluating the Kubo formula for the Hamiltonian Eq.~(\ref{2bandH})~\cite{qi:06}. The Kubo formula Eq.~(\ref{Kubo}) 
can be cast into the form:
\beq
C_\t{TKNN}=\frac{1}{2\pi}\int_{BZ}\!\rd^2 k\,\Omega_\t{B}({\bk})
\label{TKNN-Berry}
\eeq
where $\Omega_\t{B}({\bk})$ is the Berry curvature~\cite{thouless:82}
\begin{align}
\Omega_\t B({\bk})&=i\sum_n\bm\nabla_{\bk}\times\langle n,{\bk}|\bm\nabla_{\bk}|n,{\bk}\rangle|_z\,.
\label{BC}
\end{align}
For the two-band model system we identify $|\pm,{\bk}\rangle$ with $\bu_{\pm}(\bk)$ from Eq.~(\ref{spinor}) to arrive at Eq.~\eqref{berryvec}.

To summarize, we  identified $2 \Om_\t{B}(\bk)$ 
with the integrand of the skyrmion number integral, either in terms of the normalized Bloch vector $\Om_{\rm h}(\bk)$
or in terms of the normalized spin vector $\Om_{\rm S}(\bk)$
\beq
2\,\Om(\bk)= \Om_{\rm h}(\bk)=\Om_{\rm S}(\bk) =2\,\Omega_\t{B}(\bk).
\eeq
The skyrmion number integral $N_{\rm S}$ is the Brouwer degree $C_{\rm Brouwer}$ of the 
map $\bh(\bk)/h(\bk)$ from the torus to the unit sphere
\beq
\frac{1}{2\pi}\int_{BZ}\bOm = C_{\rm Brouwer}= N_{\rm S} = C_\t{TKNN},
\eeq
which equals the TKNN number $C_\t{TKNN}$ of the transverse Hall conductance.

The topological insulator is distinct from the standard band insulators through finite, integer numbered topological invariants, an example of which is the integer $C_\t{TKNN}$. 
This is true for electronic systems with discrete translational invariance and, consequently, a well-defined BZ. 
The general mechanism to convert a band insulator into a topological insulator is band inversion; spin-orbit coupling
is the underlying mechanism which inverts the usual ordering of conduction and valence bands~\cite{bernevig:06}.
In the example for a skyrmion texture of the Bloch vector in Fig.~\ref{Blochvector} band inversion is easily achieved
by sign change to a negative parameter $M$ (see caption of Fig.~\ref{Blochvector})~\cite{qi:06}.

\section{S-Wave Superconductor with Spin-Orbit Coupling}
\label{sec:topo}

\begin{figure}[t!]
\centering
\vspace{3mm}
\begin{overpic}
[width=0.98\columnwidth]{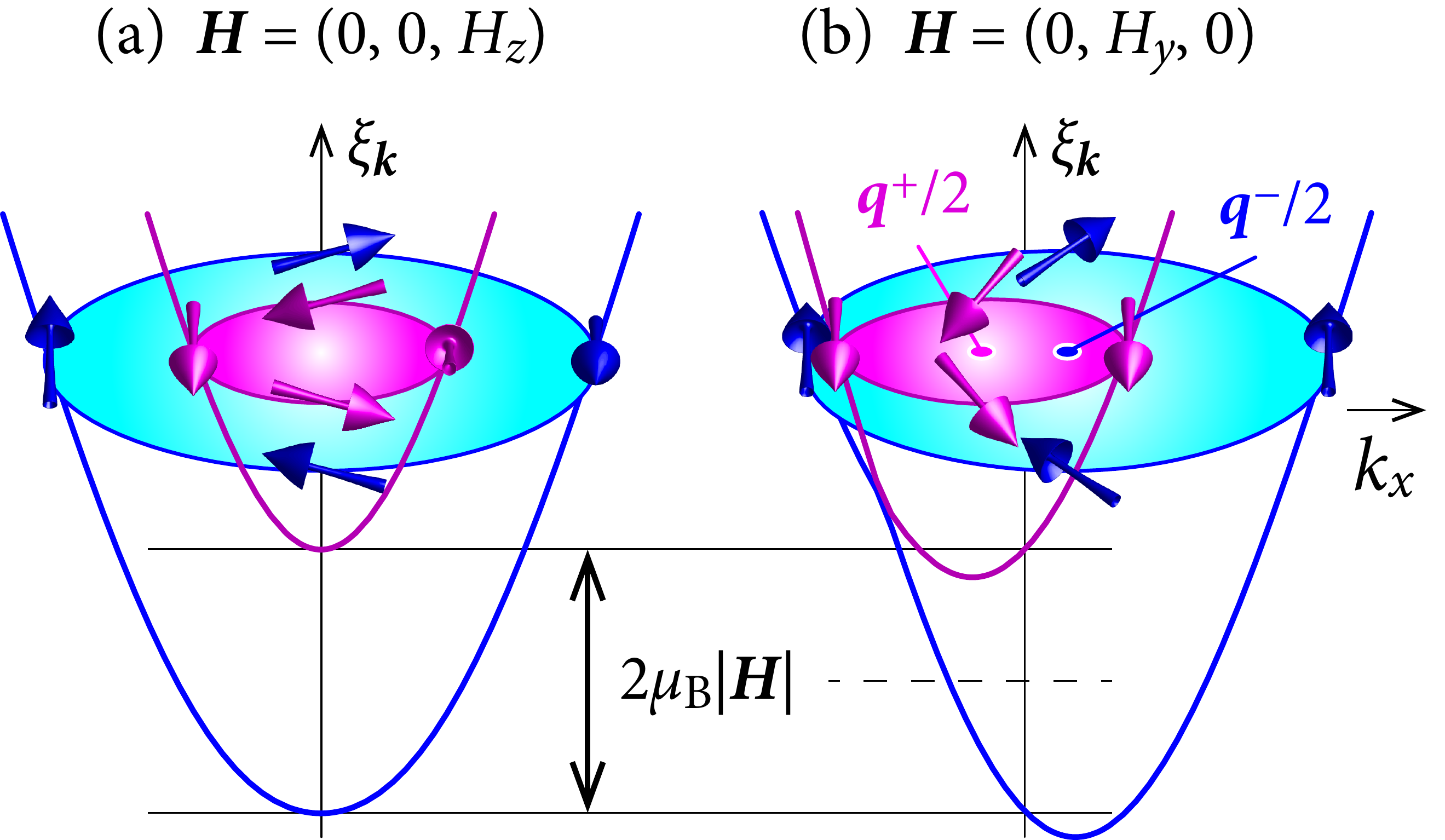}
\end{overpic}
\vspace{-0mm}
\caption{(Color online) Band dispersions $\xi^+_{\bk}$ (pink), $\xi^-_{\bk}$ (blue) with Rashba 
spin-orbit coupling and (a) out-of-plane and (b) in-plane Zeeman field. The centers of the shifted
Fermi surfaces in (b) are at the momenta ${\bq}^+/2=(q^+/2,0)$ and ${\bq}^-/2=(q^-/2,0)$, 
respectively, with $q^+\sim q^-$~\cite{loder:13,loder15}. 
}
\label{FermiSurfaceChirality}
\end{figure}

We now apply these concepts and the established topological invariants to a superconductor.
Specifically, we examine the topological character of a two-dimensional (2D) $s$-wave 
superconductor with Rashba SOC in an external magnetic field focussing on its spin texture in 
momentum space. The magnetic field is assumed to couple exclusively to the electron spin via the 
Zeeman energy and not to the orbital motion of the electrons. The reasoning for this ansatz is that
the topological character of the superconductor is preserved upon rotating the Zeeman field from an 
orientation perpendicular to the superconducting plane towards an in-plane orientation~\cite{loder15}. 
For the latter orientation complications due to vortex physics and circulating supercurrents are absent.

For simplicity we use a one-band tight-binding model on a square lattice with periodic boundary 
conditions and nearest-neighbor hopping amplitude $t$. We start from a BCS pairing Hamiltonian 
for a superconducting state with $s$-wave symmetry and supplement it by ${\cal H}_S$ which contains 
Rashba SOC and the Zeeman energy. We follow the general notation of the previous chapter, and with 
\begin{align}
{\cal H}_{\rm S}=\sum_{\bk;s,s'=\pm 1}(\alpha\, {\bm g}(\bk)+\mu_{\rm B}\bH)\cdot\bm\sigma_{ss'}\,c^\dag_{\bk,s}c_{\bk,s'}
\label{g1}
\end{align}
the Bloch vector is now specified as $\bh(\bk)=\alpha\, {\bm g}(\bk)+\mu_{\rm B}\bH$. Rashba SOC with strength 
$\alpha$ is represented by the vector ${\bm g}(\k)=(\sin k_y, -\sin k_x,0)$, ${\bH}$ denotes the 
external magnetic field and $\mu_{\rm B}$ is the Bohr magneton. 
Diagonalizing ${\cal H}_0+{\cal H}_S$, where ${\cal H}_0=\sum_{\bk,s}\epsilon_{\bk} 
c^\dag_{\bk, s}c_{\bk, s}$ is the kinetic energy of the hopping motion, leads to the two spin-split chiral 
energy bands $\xi^\pm_{\bk}=\epsilon_{\bk}\pm|\alpha\, {\bm g}(\bk)+\bH|$. The chemical potential $\mu$ is hereby 
combined with the dispersion as $\epsilon_{\bk}=-2t(\cos k_x+\cos k_y)-\mu$.

The two chiral bands $\xi^\pm_{\bk}$ are depicted in Fig.~\ref{FermiSurfaceChirality}. The electrons 
are supposed to move in the $x$-$y$--plane and the dispersion is shown for the two qualitatively 
different cases of an out-of-plane magnetic field ${\bH}=(0,0,H_z)$ and an in-plane magnetic field 
${\bH}=(0,H_y,0)$. In the chiral bands the spin is either parallel or antiparallel to 
${\bh}({\bk})$ and has a component which rotates either counter-clockwise or clockwise upon 
circulating the Fermi surfaces. For an in-plane magnetic field the centers of the Fermi surfaces for 
the $\xi^+_{\bk}$ and the $\xi^-_{\bk}$ bands shift in opposite directions away from the $\Gamma$ 
point and perpendicular to the magnetic field. For a finite in-plane field component the 
superconducting state will therefore necessarily involve Cooper pairs with finite center-of-mass
momenta (COMM)~\cite{larkin64,fulde64,barzykin02,kaur05,michaeli12,xu:15,loder15}. 

Allowing for an arbitrary magnetic field orientation the ansatz for the superconducting state 
has to include the option to form electron pairs with finite COMM. Assuming a local, on-site pairing 
interaction, the superconducting order parameter is calculated self-consistently from 
${\cal H}={\cal H}_0+{\cal H}_S+{\cal H}_I$ with~\cite{loder:10}
\begin{align}
{\cal H}_{\rm I}=\sum_{\bk,\bq}\sum_ss\left[\Delta_{\bq}^{\!*}c_{-\bk+\bq,-s}c_{\bk,s}+\Delta_{\bq} 
c^\dag_{\bk,s}c^\dag_{-\bk+\bq,-s}\right],
\label{g3}
\end{align}
where the summation is performed over all possible COMMs $\bq$ of the electron pairs. The singlet order 
parameter for COMM $\bq$ is calculated as
\begin{align}
\Delta_{\bq}=-\frac{V}{2N}\sum_{\bk'}\langle c_{-{\bk}'+{\bq},\downarrow}c_{{\bk}',\uparrow}-
c_{-{\bk}'+{\bq},\uparrow} c_{{\bk}',\downarrow}\rangle,
\label{selfcondelta}
\end{align}
where $V$ is the pairing-interaction strength.

If all electron pairs carry the same COMM $\bq$, ${\cal H}=\sum_{\bk}\b C^\dag_{\bk} {\cal H}({\bk})\b C_{\bk}$ is represented 
by the 4$\times$4 matrix 
\begin{align}
{\cal H}({\bk})=\begin{pmatrix}\epsilon_{\bk}\,\sigma^0+\bh({\bk})\cdot\bm\sigma & i\sigma^y\Delta_{\bq} \cr 
-i\sigma^y\Delta^{\!*}_{\bq} & -\epsilon_{-{\bk}+{\bq}}\,\sigma^0-\bh({-{\bk}+{\bq}})
\cdot\bm\sigma^*\end{pmatrix}
\label{g5}
\end{align}
with $\b C^\top_\k=(c_{{\bk},\uparrow},c_{{\bk},\downarrow},c^\dag_{-{\bk}+{\bq},\uparrow},
c^\dag_{-{\bk}+{\bq},\downarrow})$.

It is instructive to rewrite the Hamiltonian in the {\it helicity basis}, i.e. in terms of the 
quasiparticle operators $a_{{\bk},\pm}$ which diagonalize ${\cal H}_0+{\cal H}_S$ \cite{SigristAIP}. 
These operators generate the band eigenstates with energies $\xi_{\bk}^\pm$ and are obtained by the 
transformation
\begin{equation}
{a_{{\bk},+}\choose a_{{\bk},-}}=\displaystyle{1\over\sqrt{2h({\bk})}}\left(\begin{array}{c@{\quad}c} 
\sqrt{h({\bk})+h_z({\bk})} & \phi_{\bk}^+ \\ -\sqrt{h({\bk})-h_z({\bk})}  & 
\phi_{\bk}^- \end{array}\right){c_{{\bk},\uparrow}\choose c_{{\bk},\downarrow}}
\end{equation}
with $h({\bk})=|{\bh}({\bk})|$ and 
\begin{equation}
\phi_{\bk}^\pm =\displaystyle{h_x({\bk})- i h_y({\bk})\over \sqrt{h({\bk})\pm h_z({\bk})}} \, .
\end{equation}
In the helicity basis, for a single COMM ${\bq}$, the Hamiltonian takes the form ${\cal H}=
\sum_{\bk}\b {\cal A}^\dag_{\bk} {\cal H}^{\rm hb}({\bk})\b {\cal A}_{\bk}$ with 
$\b {\cal A}^\top_{\bk}=(a_{{\bk},+},a^\dagger_{-{\bk}+{\bq},+},a_{{\bk},-},a^\dag_{-{\bk}+{\bq},-})$ and 
the 4$\times$4 matrix
\begin{widetext}
\beq
{\cal H}^{\rm hb}({\bk})=\left(\begin{array}{c@{\quad}c@{\quad}c@{\quad}c} \epsilon_{\bk}+h({\bk}) & 
\Delta^{++}({\bk},{\bq}) & 0 & \Delta^{+-}({\bk},{\bq}) \\ {\Delta^{++}}^{\!*}({\bk},{\bq}) & 
-\epsilon_{-{\bk}+{\bq}}-h({-{\bk}+{\bq}}) & \Delta^{-+}({\bk},{\bq}) & 0 \\ 0 & 
{\Delta^{-+}}^{\!*}({\bk},{\bq}) & \epsilon_{\bk}-h({\bk}) & \Delta^{--}({\bk},{\bq}) \\ 
{\Delta^{+-}}^{\!*}({\bk},{\bq}) & 0 & {\Delta^{--}}^{\!*}({\bk},{\bq}) & 
-\epsilon_{-{\bk}+{\bq}}+h({-{\bk}+{\bq}}) \end{array} 
\right) \, .
\label{Hhb}
\eeq
\end{widetext}
$\Delta^{+-}({\bk},{\bq})$ and $\Delta^{-+}({\bk},{\bq})$ denote the inter-band pairing 
amplitudes. The intra-band pairing amplitudes $\Delta^{++}({\bk},{\bq})$ and $\Delta^{--}({\bk},
{\bq})$ are odd functions with respect to interchanging the momenta ${\bk}$ and 
$-{\bk}+{\bq}$~\cite{loder:13}, as can be explicitly verified from their functional form given  
in Appendix~\ref{app0}. The matrix elements on the skew diagonal in Eq.~(\ref{Hhb}) lead to structures 
in the density of states away from the Fermi energy. Hence, they do not affect the closing of the bulk 
energy gap and an eventual topological phase transition~\cite{koenig:08}.

For the special case $H_x=H_y=0$, i.e. for an out-of-plane orientation of the
magnetic field, intra-band pairing does not require a finite COMM, and for ${\bq}={\bf 0}$ the 
pairing amplitude simply reduces to 
\begin{eqnarray}
\Delta^{++}({\bk},{\bf 0}) &=& -{\Delta_{\bf 0}\over h({\bq})}\left( h_x({\bk})-i
h_y({\bk})\right) \nonumber \\ 
&=& -g{\Delta_{\bf 0}\over h({\bk})}\left( \sin k_y + i\sin k_x \right) \, ,
\end{eqnarray} 
which is evidently odd in ${\bk}$. In fact, the pairing state in each of the helical bands 
acquires the form of a spinless $p_y +ip_x$ superconductor, similar to the proximity induced 
superconductivity on the surface of a topological insulator in contact with an $s$-wave 
superconductor~\cite{FuKane08} or in a semiconductor quantum well coupled to an $s$-wave 
superconductor and a ferromagnetic insulator~\cite{alicea:10,sau:10}. The order parameters on the 
two Fermi surface sheets are equal in magnitude, but since 
\begin{equation}
\Delta^{--}({\bk},{\bf 0})=g{\Delta_{\bf 0}\over h({\bk})}\left( \sin k_y -i\sin k_x
\right) \, ,
\end{equation}
they have the opposite chirality~\cite{alicea:10}. Beyond a critical magnetic field strength $H_{\rm t}$, the upper 
$\xi_{\bk}^+$ band is unoccupied; $\Delta^{++}$ and the inter-band pairing amplitudes then
necessarily vanish. The pairing on the remaining lower $\xi_{\bk}^-$ band has a unique chirality. 
This is the origin for a topologically non-trivial character of the superconducting state for $H>H_{\rm t}$.

\begin{figure}[t!]
\begin{overpic}
[width=0.95\columnwidth]{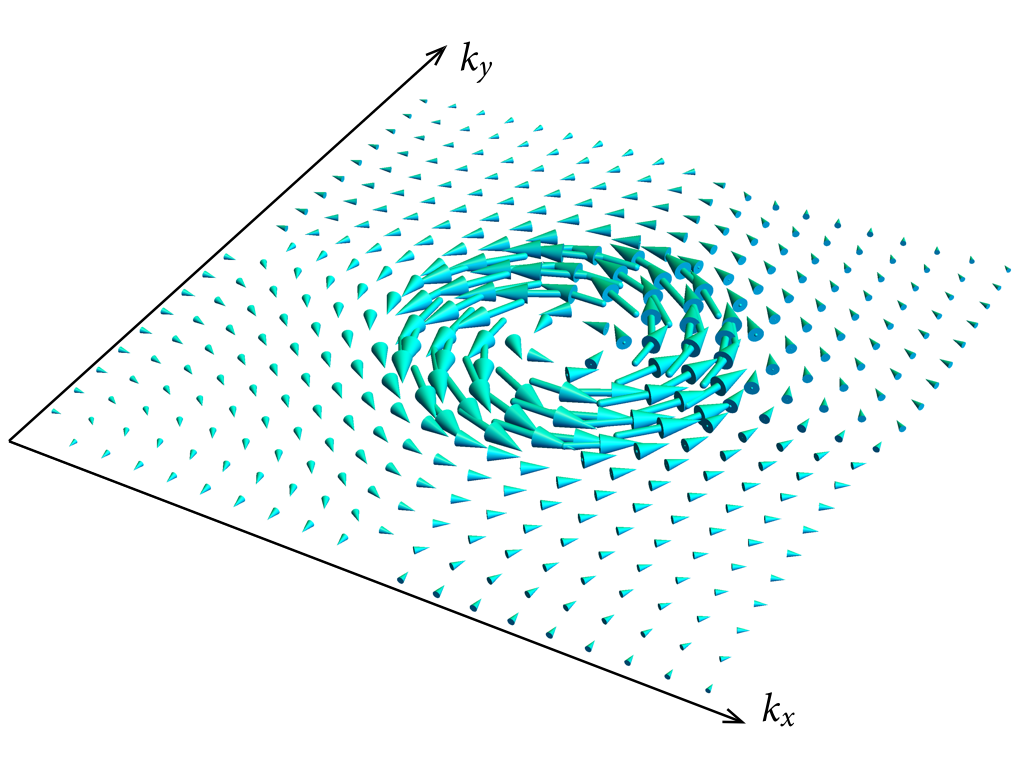}
\put(0,66){a) $H_z=0$}
\end{overpic}\\
\begin{overpic}
[width=0.95\columnwidth]{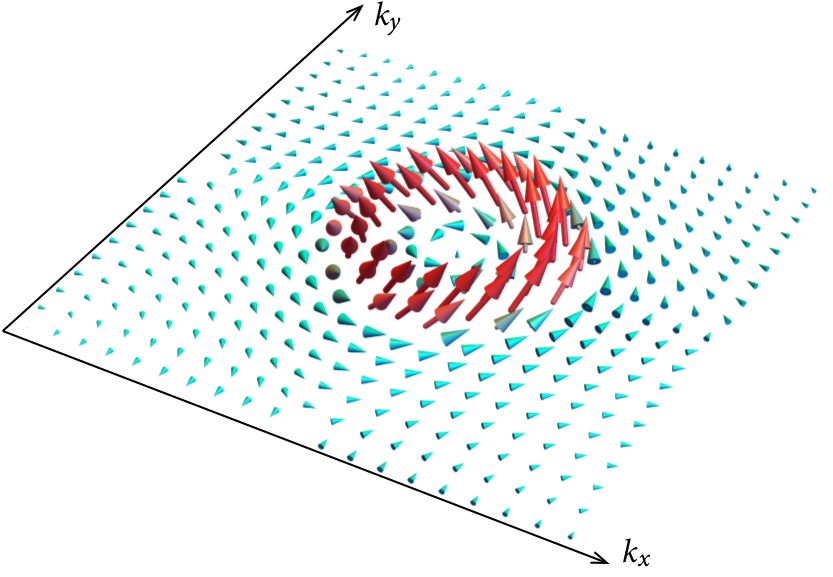}
\put(0,66){b) $\mu_{\rm B} H_z=0.2\,t$}
\end{overpic}\\
\begin{overpic}
[width=0.95\columnwidth]{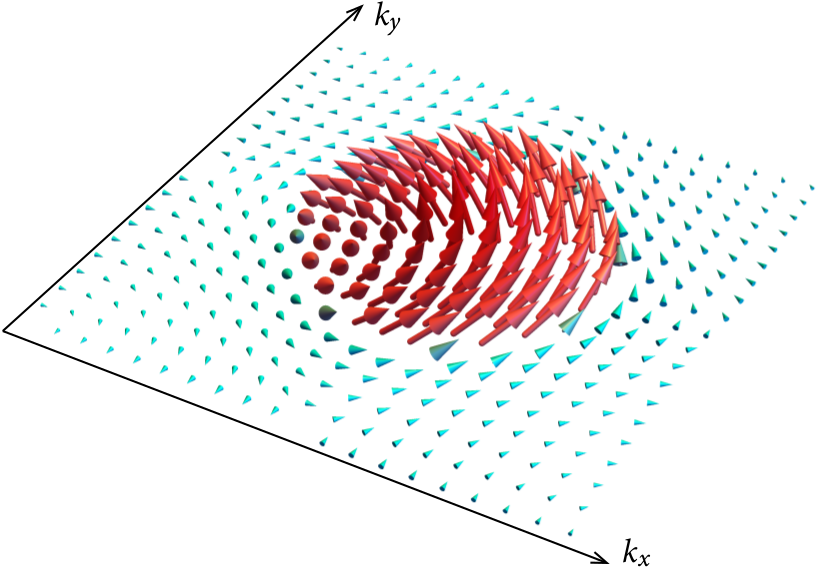}
\put(0,66){c) $\mu_{\rm B}H_z=0.5\,t$}
\end{overpic}\\
\vspace{3mm}
\begin{overpic}
[width=0.6\columnwidth]{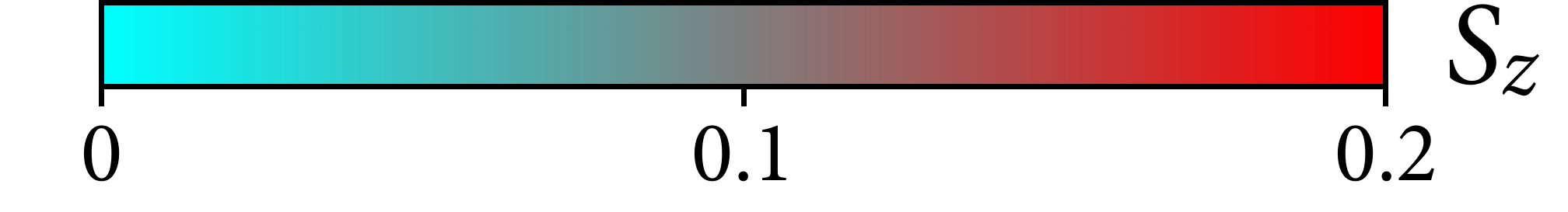}
\end{overpic}
\vspace{-0mm}
\caption{(Color online) Ground-state spin-expectation value ${\bS}({\bk})$ in
 the first Brillouin zone, $k_x,k_y\in[-\pi,\pi]$, for an out-of-plane orientation of
the magnetic field, i.e. $H_x=H_y=0$. The further parameters are: $\Delta=0.15\,t$, $\alpha=0.4\,t$, 
and $\mu=-3.2\,t$. The critical Zeeman field for these parameters is $\mu_{\rm B}H_{\rm t}\sim 0.4t$. The color of 
each spin represents the value of the out-of-plane component $S_z({\bk})$; its magnitude is 
determined from the color bar at the bottom. The skyrmion counting number is $N_{\rm S}=0$ in (a) and (b), 
while $N_{\rm S}=1/2$ in (c).
}
\label{Fig3}
\end{figure}

\section{Momentum-Space Spin Textures}
\label{sec:textures}

The superconducting state is represented by the 4$\times$4 matrix of Eq.~(\ref{g5}) in Nambu 
space. The four eigenvectors 
$\bu_n(\bk)$ describe 
the specific mixture of electrons and holes in each eigenstate for a given momentum
${\bk}$. In the general situation with SOC and an arbitrarily oriented magnetic  
${\cal H}({\bk})$ cannot be written in a block-diagonal form due to the interband pairing amplitudes, as is 
evident from the structure of the Hamiltonian matrix ${\cal H}^{\rm hb}({\bk})$ (see Eq.~(\ref{Hhb})) in the 
helicity basis. In this general case, the eigenvectors $\bu_n(\bk)$ are therefore calculated 
numerically which thereby requires to determine the optimum COMM ${\bq}$. The eigenvectors are 
subsequently used to evaluate the spin expectation values ${\bS}({\bk})$ in the 
superconducting state. Their texture in momentum space is analyzed separately for an 
out-of-plane orientation of the magnetic field and for a mixed situation, in which the magnetic 
field has both out-of-plane and in-plane components. 

\subsection{Out-of-plane magnetic field, $\bm{T=0}$}
\label{sec:Hztextures}

We start with a magnetic field ${\bH}=(0,0,H_z)$, for which the two Fermi surface sheets are 
both centered at the $\Gamma$-point and intra-band electron pairs form with zero COMM. In this
comparatively simple situation we refrain from a self-consistent calculation and assume a finite
fixed order parameter $\Delta\equiv\Delta_{{\bq}={\bf 0}}$. Figure~\ref{Fig3} illustrates the spin 
texture in the first BZ for three different field strengths, including $H_z=0$. These 
results were obtained at zero temperature for the parameter set given in the caption. 
Figures~\ref{Fig3}a and ~\ref{Fig3}b depict the texture in the topologically trivial 
superconducting state. For $H_z=0$, there is no net spin magnetization, and ${\bS}({\bk})$ is 
confined to the $x$-$y$--plane. The spin winding reflects the spin-momentum 
locking due to SOC. For the finite Zeeman field in Fig.~\ref{Fig3}b, imprints of the two 
$\Gamma$-point centered Fermi surfaces with unequal areas are visible, and the electrons as a 
whole have picked up a finite spin magnetization. For the larger Zeeman field in Fig.~\ref{Fig3}c, 
$H_z$ has exceeded its critical value $H_{\rm t}=\sqrt{\Delta^2+\epsilon_{\bf 0}^2}$~\cite{sato2:09}. For 
$H_z > H_{\rm t}$, all electrons are
in the lower $\xi_{\bk}^-$ band and 
form a topological superconductor.  

\begin{figure*}[t]
\includegraphics[width=1.99\columnwidth]{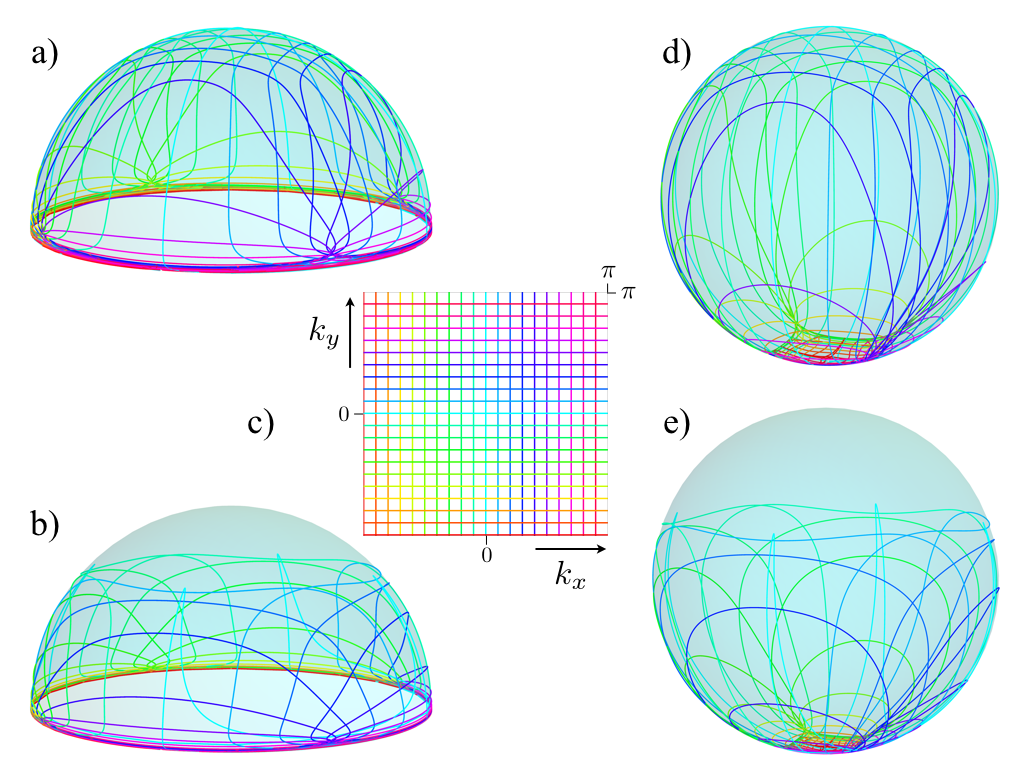}
\caption{(Color online) The maps $\hat{\bS}(\bk)$ [(a) and (b)]and $\hat{\bS}_{\rm reg}(\bk)$ [(d) and (e)] for the topologically 
non-trivial [(a) and (d)] and trivial cases [(b) and (e)].
The colored rectangular grid in the BZ, depicted in (c), is mapped onto the corresponding 
lines on the (hemi-)spheres.  Note the approximate restoration of the rectangular grid 
structure in the vicinity of the ``south pole'' in (d) and (e). For $N_{\rm S}^{({\rm meron})}\!\!=\!1$ [(a) and (d)] 
an almost regular pattern is also visible close to the ``north pole''. The distortion of the flat 
metric in the BZ is only weak in these areas. For these maps, the lattice size is $1600 \times 1600$,
the chemical potential is $\mu\! =\! -3.6\, t$ (which corresponds to a density $n = 0.075$), the Rashba SOC 
strength is $\alpha = 0.6\, t$, and the gap is $\Delta\!= 0.15\, t$. The magnetic field is $\mu_B H_z\! =\! 0.4\, t$ for the
trivial cases [(b) and (e)] and $\mu_B H_z\! =\!  t$ for the topologically non-trival cases [(a) and (d)].}
\label{hemispheremap}
\end{figure*}

The topological transition is captured by the changes in the spin texture~\cite{annica:14,dong:15}.
In order to verify that the spin texture indeed signifies the transition, we calculate the skyrmion
number as the corresponding 
topological invariant. For this purpose, we can utilize the normalized
spin-vector expectation value ${\hat {\bS}}({\bk})={\bS}({\bk})/|{\bS}({\bk})|$ for the required BZ
integral, as outlined in chapter \ref{sec:concepts}. 
With the spin product
\begin{align}
\Omega_{\rm S}({\bk})={\hat {\bS}}({\bk})\cdot\left(\partial_{k_x}{\hat {\bS}}({\bk})\times
\partial_{k_y}{\hat {\bS}}({\bk})\right) 
\label{c7.3}
\end{align}
the skyrmion number of the spin texture is 
\begin{equation}
N_{\rm S} = \frac{1}{4\pi}\int_{BZ}\Omega_{{\rm S}}({\bk}){\rm d}^2{k} \, .
\label{nsintegrals}
\end{equation}
$N_{\rm S}$ takes integer values if the map ${\hat {\bS}}({\bk})$ from the BZ to the unit sphere is surjective and everywhere continuous.

We now introduce a different normalization of the expectation value of the spin, ${\bS}_{\rm L}({\bk})$, in which 
$S_x$ and $S_y$ are normalized only to the in-plane components
\begin{align}
{\bS}_{\rm L}({\bk})=\left(\frac{S_x}{\sqrt{S_x^2+S_y^2}}, \frac{S_y}{\sqrt{S_x^2+S_y^2}}, 
\,{S_z}\right) 
\label{c8}
\end{align}
and we define 
\begin{align}
\Omega_{\rm L}({\bk})={{\bS}_{\rm L}}({\bk})\cdot\bigl(\partial_{k_x}{{\bS}_{\rm L}}({\bk})\times
\partial_{k_y}{{\bS}_{\rm L}}({\bk})\bigr) 
\label{c7.4}
\end{align}
This spin product has the remarkable property to be pointwise identical to the Berry curvature:
\begin{equation}
\Omega_{{\rm L}}({\bk})= \Omega_{\rm B}({\bk})
\label{Omegabar_equal_OmegaBerry}
\end{equation}
for each momentum ${\bk}$. Heretofore, this equality could be confirmed only numerically.
The zero-temperature quantity $\Omega_{\rm L}$ will prove useful when we extend its range of validity to finite 
temperatures in Sec.~\ref{sec:outofplaneT}. 

In the absence of an analytical proof of the 
equality~(\ref{Omegabar_equal_OmegaBerry}) we resort to the relations established for the two-band model
in Sec.~\ref{app1}: In Eq.~(\ref{Sbar_1}) we introduced a similar spin quantity, $\bar {\bS}_1({\bk})$,
where the particular normalization of $\bar {\bS}_1({\bk})$ describes the mapping of the BZ 
to the barrel of a cylinder with unit radius---while
that of ${\hat {\bS}}({\bk})$ involves the projection onto the unit sphere. Only ${\hat {\bS}}({\bk})$
and $\bar {\bS}_1({\bk})$ are proven to generate the topological invariant $N_{\rm S}$. 
Nevertheless, it is
striking that Eq.~(\ref{Omegabar_equal_OmegaBerry}) holds. Moreover, $2N_{\rm S}$ of Eq.~(\ref{nsintegrals})
is identical to the Chern (or TKNN number) given in Eq.~(\ref{TKNN-Berry})
\begin{equation}
2N_{\rm S} = \frac{1}{2\pi}\int_{BZ}\Omega_{{\rm B}}({\bk}){\rm d}^2{k}
\label{nsaddintegrals}
\end{equation}
in the present case. While the relation between the skyrmion number of the spin texture  as given in
Eq.~(\ref{nsintegrals}) and the Berry curvature is a numerical finding, the spin texture may be analyzed analytically in its own right, as is done in the following.

For the spin textures in Figs.~\ref{Fig3}a and \ref{Fig3}b with the Zeeman fields 
$H_z < H_{\rm t}$ the BZ integrals lead to $N_{\rm S} =0$. Instead, for the 
topological superconductor with the Zeeman field $H_z > H_{\rm t}$ as in Fig.~\ref{Fig3}c, we obtain 
$N_{\rm S}=1/2$. 
This latter non-trivial result may appear surprising at first sight, because 
$N_{\rm S}$ must be always zero, if the spin texture is a continuous map from the BZ to the 
unit sphere $S^2$. Here, in fact this map can never cover the entire sphere because $S_z$ 
attains only non-negative values for $H_z\ge 0$. However, the topological superconductor 
the spin texture corresponds to a very special {\em singular} map from the BZ to the upper 
half-sphere, which preserves a topological signature, as shown below. 

Figure~\ref{Fig3} illustrates that the components of ${\bS}({\bk})$ become very small at the 
BZ boundary. In fact, ${\bS}({\bk})$ vanishes at eight (in part equivalent) BZ boundary points
${\bk}_c$, namely at the BZ corners $(\pm\pi,\pm\pi)$ and at the four points $(\pm\pi,0)$ and 
$(0,\pm\pi)$ on the BZ faces. While also the normalized component 
$S_z({\bk})/|{\bS}({\bk})|$ vanishes at these ${\bk}_c$, either
$\lim_{{\bk}\rightarrow{\bk}_c} S_x({\bk})/|{\bS}({\bk})|$ or $\lim_{{\bk}\rightarrow{\bk}_c} S_y({\bk})/|{\bS}({\bk})|$ remain finite.  
The normalized vector field $\hat{{\bS}}({\bk})$ can  be 
defined at all  points ${\bk}\neq{\bk}_c$ and the
integrand in Eq.~\eqref{nsintegrals} is therefore well defined in the BZ, away from the momenta $\bk_c$ 
on the BZ boundary. Considering the BZ as a chart of the torus $T^2$, the eight points 
$\bk_c$ correspond to three points on the torus: 
 $(\pm\pi,\pm\pi)$ are equivalent as well as $(\pm\pi,0)$ and $(0,\pm\pi)$. For $H_z<H_{\rm t}$, 
another point with the same properties appears at the center of the BZ, $\bk_c={\bf 0}$. 

Close to these
three (respectively four) points on the torus, the spin texture exhibits ``vortex-like'' characteristics.
Due to such micro-vortices in momentum-space, the vector field $\hat{{\bS}}$ 
is discontinuous at the momenta ${\bk}_c$ (and also at ${\bk_c =\bf 0}$ for $H_z<H_{\rm t}$) 
because the image of each ${\bk}_c$ is the full equator ($S_z=0$) of the unit sphere $S^2$. 
More precisely, an infinitesimally small 
circle ${\bk}_c+\varepsilon\,(\cos\phi\, {\bm{e}}_1+\sin\phi\, {\bm{e}}_2)$, with $\varepsilon\ll 1$ 
and two orthogonal unit vectors ${\bm{e}}_1$ and ${\bm{e}}_2$, is mapped onto 
$\big(\sqrt{1-\tilde{\varepsilon}^2}\cos\th(\p), \sqrt{1-\tilde{\varepsilon}^2}\sin\th(\p),\tilde{\varepsilon}(\p)\big)$, where $\th(\p)$ and $\tilde{\varepsilon}(\p)$ are functions of $\p$
and
$\tilde{\varepsilon}(\p)$ is of the order $\varepsilon$.  
As the map $T^2 \rightarrow S^2$  furnished by $\hat{{\bS}}({\bk})$ is {\it not continuous} at the 
isolated points ${\bk}_c$, no topologically meaningful quantity appears to be associated with it and the evaluation of 
Eq.~\eqref{nsintegrals} may be expected to yield arbitrary values.

Yet, the skyrmion counting number $N_{\rm S}$ is quantized, albeit not in integers but in half-integers. 
This apparent puzzle is resolved by the following observation:  
$S_z$ is confined to values $\ge 0$ for $H_z\ge 0$, signifying that the BZ 
is mapped into the upper hemisphere $S^2_u$. 
This manifold has a boundary $A$ (the equator) and is topologically equivalent to the disk $D^2$.
The map given by $\hat{{\bS}}({\bk})$ is continuous from $\{{\bk} \in \textrm{BZ}, {\bk}\neq{\bk}_c \}$ to the open hemisphere $S^2_u \setminus A$.
Because the equator is the unique limit set of the images of any sequence of points converging 
towards one of the isolated 
points ${\bk}_c$,  
we can construct from  $\hat{{\bS}}$ a {\em continuous} 
map $\hat{\bS}_{\rm reg}= {\bf \Phi}\circ\hat{{\bS}}$ of the torus 
onto $S^2$, a compact manifold without boundary, by compactifying the open hemisphere 
$S^2_u\setminus A$ to a sphere via the 
function ${\bf \Phi}$, with
\begin{align}
{\bf \Phi}(x,y,z)=\Biggl(2x\sqrt{\frac{z}{z+1}},\; 2y\sqrt{\frac{z}{z+1}},\; 2z-1\Biggr)\, .
\end{align}
Here $\bf \Phi$ maps all points located on the equator of $S^2_u$ to the south pole $(0,0,-1)$ of $S^2$. 
 In turn, $\hat{\bS}_{\rm reg}({\bk})$ maps the
  singular points ${\bk}_c$ to  $(0,0,-1)$, a neighborhood of any ${\bk}_c$ 
  to a neighborhood of the south pole, and $\hat{\bS}_{\rm reg}({\bk})$ is therefore continuous. The application of ${\bf \Phi}$ multiplies the surface element of $S^2$ by 2 (compare Eq.~\eqref{brouwer2}). Consequently, we define the quantity
\begin{equation}
N^{({\rm meron})}_{\rm S} = \frac{1}{2\pi}\int_{BZ}\Omega_{ {\rm S}}({\bk}){\rm d}^2{k}=
\frac{1}{4\pi}\int_{BZ}\Omega_{\rm reg}({\bk}){\rm d}^2{k}\, .
\label{meron}
\end{equation}
which always takes integer values. In analogy to the relations (\ref{c7.3}) and (\ref{c7.4}) we have introduced  the spin product
\begin{align}
\Omega_{\rm reg}({\bk})={\hat{\bS}_{\rm reg}}({\bk})\cdot\left(\partial_{k_x}{\hat{\bS}_{\rm reg}}({\bk})\times
\partial_{k_y}{\hat{\bS}_{\rm reg}}({\bk})\right)\, . 
\label{c7.5}
\end{align}
We conclude that $N_{\rm S}$ is quantized in half-integers, and we associate the momentum-space spin structures 
in Fig.~\ref{Fig3} with {\it merons} 
(see, e.g. Ref.~\onlinecite{volovikbook}), and  $N^{({\rm meron})}_{\rm S}$ is their counting number.

\begin{figure}[h]
\includegraphics[width=0.73\columnwidth]{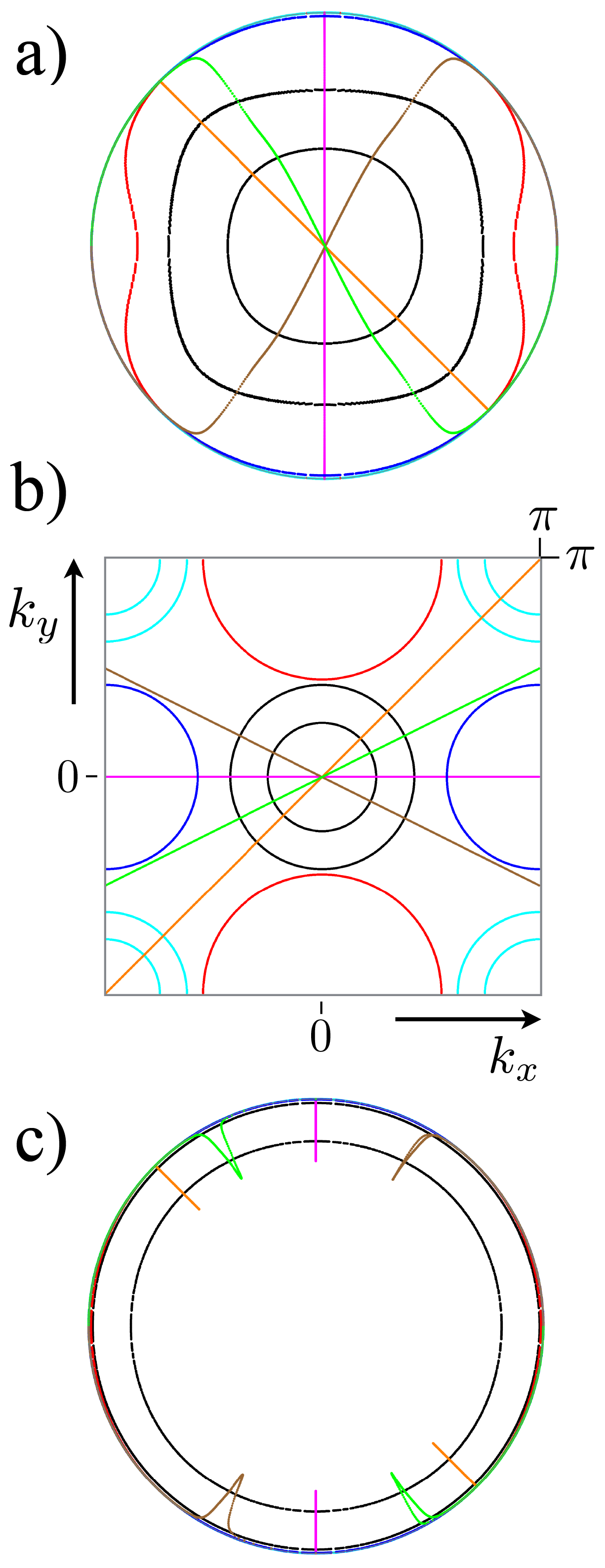}
\vskip0.3truecm
\caption{(Color online) Images of selected paths in the BZ as obtained 
from the map $\hat{{\bS}}(\bk)$; the paths are shown in panel (b).
The topologically non-trivial case for $H_z>H_{\rm t}$ is presented in (a) and the topologically trivial case in (c). 
The image points
in the upper hemisphere are mapped by vertical projection [$(x,y,z)\rightarrow(x,y,0)$] onto the unit disk. $S^2_u$ is completely covered for $H_z>H_{\rm t}$  while for $H_z<H_{\rm t}$ a 
part of $S^2_u$ is covered twice, and the overall coverage is incomplete. 
The parameter sets are the same as in Fig.~\ref{hemispheremap}. 
}
\label{Fig5}
\end{figure}

\begin{figure*}[t]
\includegraphics[width=1.99\columnwidth]{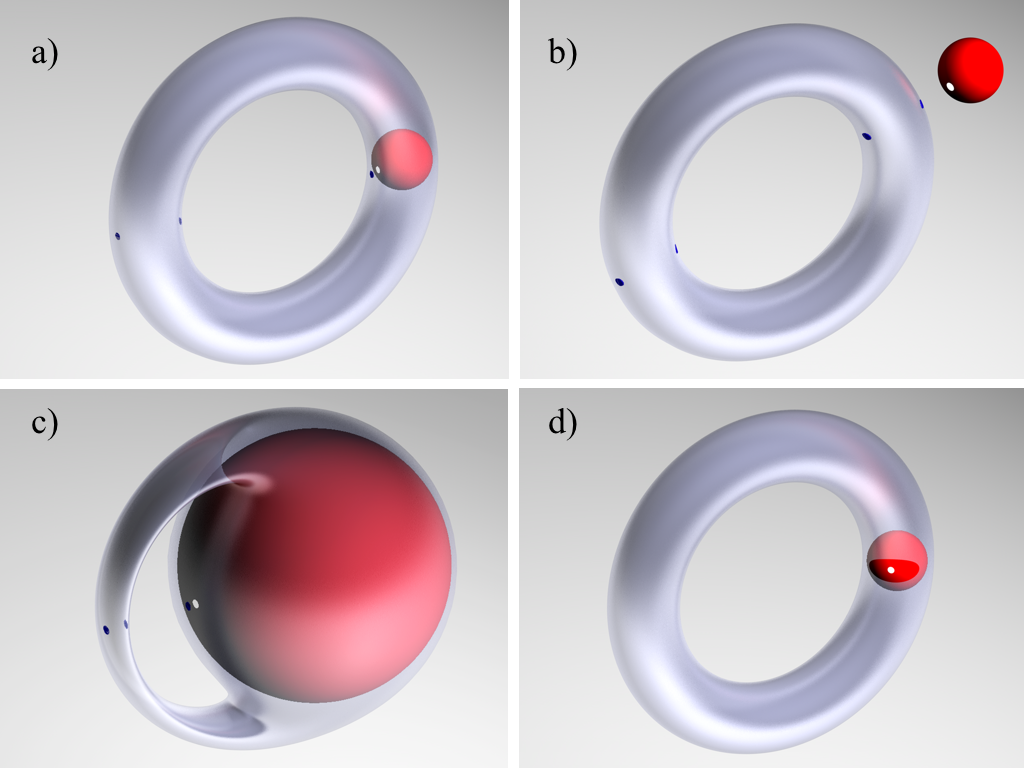}
\caption{(Color online) Visualization of the map $\cal M$ from the torus, i.e.\ the Brillouin zone, 
to the unit sphere of normalized
spin expectation values. For the topologically non-trivial state (a), the sphere is placed inside the torus and the rays from the 
center of the sphere intersect the torus either once or three times. The south pole on the sphere is indicated by a white dot,
and the ray through the south pole intersects the torus in the three blue dots. In the BZ these blue dots represent the $\bk_c$,
the centers of the three micro-vortices with vanishing $S_z$. For the topologically trivial state (b), the sphere 
is placed outside the torus and the rays from the center of the sphere intersect the torus either zero, two or four times. 
The topology of (c) is equivalent to that of (a); however the torus here is so much distorted by inflating the sphere that the multiple projection of the torus onto the sphere through $\cal M$, in the neighborhood of the south pole, is made particularly evident. In contrast, the characteristics of the map for finite temperature is sketchily represented by a sphere which pierces the torus in (d); here the intersecting torus and sphere are not related to a topological state; it is rather a candidate situation for finite temperature above a topological transition (see Sec.~\ref{sec:outofplaneT}).
}
\label{Torus_Sphere}
\end{figure*}

We can now characterize the spin textures for $H_z$ above and below $H_{\rm t}$ topologically, 
using the mapping degree of $\hat{\bS}_{\rm reg}$ in both cases. 
For $H_z>H_{\rm t}$, the complete upper hemisphere is covered by $\hat{\bS}$ 
(see Fig.~\ref{hemispheremap}a and also Fig.~\ref{Fig5}a below). The three 
points ${\bk}_c$ are mapped onto the equator and the northpole $(0,0,1)$ 
has a unique preimage. The regularized  $\hat{\bS}_{\rm reg}$ maps the BZ to the 
full sphere, the south pole of which is a regular point with three preimages. This is 
the characteristics of Fig.~\ref{hemispheremap}d. 
As the number of preimages of all points of $S^2$ is odd, the map must be topologically non-trivial.
Indeed, the Brouwer degree sums to 1 and therefore $N^{({\rm meron})}_S=1$.

The map $\hat{\bS}$ in the topologically trivial case $H_z<H_{\rm t}$ is characterized by four preimages 
of the equator (respectively the south pole for $\hat{\bS}_{\rm reg}$). 
In addition to the three points ${\bk}_c$ on the BZ boundary, the center of the BZ, ${\bk}_c={\bf 0}$, is also mapped 
to the equator while the area around the north pole $(0,0,1)$ is not covered by the maps $\hat{{\bS}}$ and 
$\hat{\bS}_{\rm reg}$. Each regular point of the map in $S^2_u$, and respectively in $S^2$, has an 
even number of preimages in the BZ, $0,2$ or $4$, and the Brouwer degree sums to 0. The 
singular map $\hat{\bS}$ is shown in  Fig.~\ref{hemispheremap}b and the regular 
map $\hat{\bS}_{\rm reg}$ in Fig.~\ref{hemispheremap}e.

Figures~\ref{Fig5}a and \ref{Fig5}c display the images of some selected paths in the BZ 
(shown in Fig.~\ref{Fig5}b) onto $S^2_u$ 
by using the vertical projection of $S^2_u$ onto the unit disk in the $x-y$-plane. The case $H_z>H_{\rm t}$ 
is represented in 
Fig.~\ref{Fig5}a. The black circles around the center of the BZ are mapped to a neighborhood of the north 
pole, while the images of the areas around the singular points ${\bk}_c$, at the BZ face centers and corners 
(blue/red/cyan circles), are located very close to the equator.

On the other hand, for $H_z<H_{\rm t}$ and $N^{({\rm meron})}_{\rm S}=0$ 
displayed in Fig.~\ref{Fig5}c, the central area of the BZ is mapped 
into the vicinity of the equator as well, while the center of the disk (the north pole) is never reached. 
We infer from the 
lines traversing the BZ (orange, magenta, brown and green), that the torus is ``folded back'' before 
reaching the north pole and does not wrap around the hemisphere. Note that for both cases,
$H_z>H_{\rm t}$ and $H_z<H_{\rm t}$, the symmetric paths, 
the diagonal (orange) line
and the horizontal (magenta) line in the BZ are mapped to straight lines in the projection of 
the hemisphere.     

These results allow to understand the topological difference between both types of maps 
(see Figs.~\ref{hemispheremap}d and \ref{hemispheremap}e) in an 
intuitive way. We can construct a map ${\cal M}$ from $T^2$ to $S^2$ as follows:
Consider a unit sphere together with a torus as subsets of $\Rr^3$ (see Fig.~\ref{Torus_Sphere}).
The rays $\lambda\bv$ from the center of the sphere---with a unit vector $\bv$ and $\lambda > 1$---intersect 
the torus at points $\lambda_1[\bv]\bv,\ldots,\lambda_j[\bv]\bv$. Then we define the map $\cal M$ from 
the torus to $S^2$ by
${\cal M}(\lambda_l[\bv]\bv)=\bv$ for all $l$. In Figs.~\ref{Torus_Sphere}a, \ref{Torus_Sphere}b and \ref{Torus_Sphere}c 
we consider a ray through the south pole of the sphere (white point) which intersects the torus either three times 
(Figs.~\ref{Torus_Sphere}a and \ref{Torus_Sphere}c) or four times (Fig.~\ref{Torus_Sphere}b)
at the blue points. In Fig.~\ref{Torus_Sphere}c, the torus is so much distorted (by ``inflating'' the sphere)
that the three sections of the torus, which are mapped onto the sphere in the neighborhood of the 
south pole, are made compellingly explicit: the rather flat areas of the ``handle'' produce the square-like grid
in the vicinity of the south pole in Fig.~\ref{hemispheremap}d.
These figures visualize that the map $\cal M$ is homotopically equivalent to 
$\hat{\bS}_{\rm reg}$. For $H_z>H_{\rm t}$ the entire sphere is located in the interior of 
the torus, while for $H_z<H_{\rm t}$ the sphere is outside the 
torus---see Figs.~\ref{Torus_Sphere}a and \ref{Torus_Sphere}b, respectively.

It has to be emphasized that the meron number, Eq.~\eqref{meron}, is a topological 
invariant only for the special class of singular maps presented here. Without the micro-vortices 
at the corners and face centers of the BZ, which are mapped to the equator, 
all spin textures with $S_z\ge 0$ 
everywhere are either topologically trivial ($N_{\rm S}=0$),
if the map $\hat{\bS}$ is continuous, or $N_{\rm S}$ takes non-quantized real values, 
depending on the details of the texture, if $\hat{\bS}$ is 
discontinuous but does not have the particular features described above. 
In the case that not the equator but some other curve in the upper hemisphere forms the limit set at the isolated 
singular points of the map $\hat{\bS}$, the surface integral in Eq.~\eqref{nsintegrals} will not be a 
multiple of $2\pi$. As will be discussed in the next section, this latter case indeed applies 
when the spin-expectation value is evaluated at finite temperatures. 

\subsection{Out-of-plane magnetic field, ${\bm T> \bm 0}$}
\label{sec:outofplaneT}

\begin{figure}[t!]
\centering
\vspace{3mm}
\begin{overpic}
[width=0.98\columnwidth]{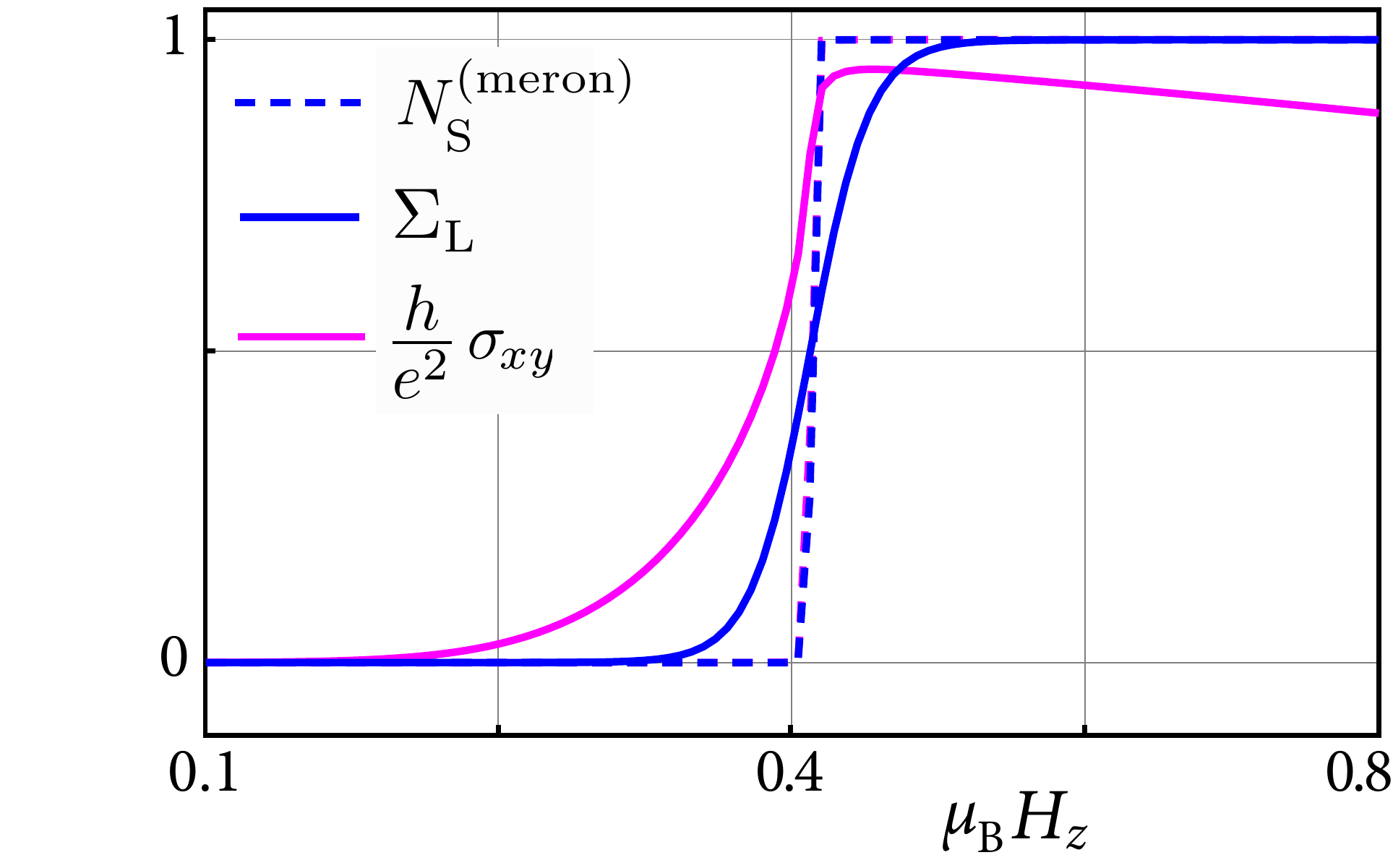}
\end{overpic}
\vspace{-0mm}
\caption{(Color online) Comparison of Zeeman-field dependences for the meron number $N^{({\rm meron})}_{\rm S} \!\!= \frac{1}{2\pi}\int_{BZ}\Omega_{ {\rm S}}({\bk}){\rm d}^2{k}$
(dashed line), the finite temperature equivalent $\Sigma_{\rm L}(T)$ (blue line), and the integral $C$ in the Kubo formula
Eq.~(\ref{Kubo}) (magenta line) for the Hall conductance. The two latter quantities are evaluated for $T=0.02\,t$.  
}
\label{fig:finiteT}
\end{figure}

For finite temperatures, $T>0$, the evaluation of the spin-expectation values requires to take 
the trace over all states and thereby also the contributions from excited states are mixed in. We first assume that the temperature is smaller than the excitation gap to paired states on the upper, unoccupied  $\xi^+_{\bk}$ band with an order parameter of the opposite chirality; for the topological 
superconductor with $H_z > H_{\rm t}$ we therefore consider the temperature regime $k_B T \ll \mu_B (H_z-H_{\rm t})$. In this case the contributions from excited states with a magnetization, which is lower than the ground-state magnetization in the external Zeeman field, are negligibly small. In essence, $S_z(T)$ is expected to increase with $T$ at these low temperatures. 
In particular, also the expectation values $S_z({\bk}_c)$ at the singular points on the BZ boundary 
turn finite and positive. The equator is therefore no longer reached by the map $\hat{\bS}(\bk)$. For the map ${\cal M}$, this situation is sketchily represented in Fig.~\ref{Torus_Sphere}d by a sphere which pierces the torus with its south pole outside. This picture visualizes that an area around the south pole no longer has preimages on the torus for the map ${\cal M}$. In a strict sense, it is therefore only the superconducting state at zero temperature that is characterized by integer valued topological invariants which are based on its spin texture in momentum space.

We define a BZ-integral over the temperature dependent skyrmion-like density 
$\Omega_{ {\rm L}\,(T>0)}({\bk})$, which is the finite temperature generalization of 
$\Omega_{ {\rm L}}({\bk})$ from Eq.~(\ref{c7.4}),  through
\begin{align} \label{Sigma_T}
\Sigma_{\rm L}(T)=\frac{1}{2\pi}\,\int_{BZ}\Omega_{ {\rm L}\,(T>0)}({\bk}){\rm d}^2{k}
\end{align}
This quantity smoothly connects to the zero temperature limit with $\Sigma_{\rm L}(T=0)=N^{({\rm meron})}_{\rm S} =2\,N_{\rm S}$ 
(cf.~Eqs.~(\ref{meron}) and (\ref{nsintegrals})).
Moreover, the Hall conductance $\sigma_{xy}(T)$ is well defined at finite temperature, 
and for $T\rightarrow 0$ the integrals $\frac{h}{e^2}\sigma_{xy}(T)$ and $\Sigma_{\rm L}(T)$
merge with the established invariants $C_{\rm TKNN} =N^{(\rm meron)}_{\rm S}$. 
The finite and zero temperature results are contrasted in Fig.~\ref{fig:finiteT}.

At zero temperature the BZ-integral over the Berry curvature, the integral over the skyrmion density, and the
Hall conductance,
all yield a step-like behavior at $H_z = H_{\rm t}$ (dashed step-function in Fig.~\ref{fig:finiteT}).
By contrast, the step is smoothed for finite temperature---however, qualitatively different for the skyrmion-like
number $\Sigma_{\rm L}(T)$ and the Hall conductance. Even though $\Sigma_{\rm L}(T)$ well approaches 
the values 0 and 1 for $H_z$ away from $H_z = H_{\rm t}$, this quantity is not a topological invariant for finite temperature
because such an invariant has to adopt these discrete values strictly. 
Yet, we emphasize that the physics at 
zero temperature is continuously approached for $T\rightarrow 0$, and the spin texture in momentum 
space maintains some of its qualitative features also at finite temperatures. This is suitably characterized by $\Sigma_{\rm L}(T)$.

Eventually, if the temperature is larger than the excitation gap to the states 
with opposite helicity ($k_B T \gtrsim \mu_B(H_z-H_{\rm t}$)), also states with 
spin $z$-component antiparallel to the applied field are thermally excited. 
However, for the relevant case of
$H_z>H_{\rm t}$, these states are still separated by a small yet finite gap 
(except precisely at $H_z=H_{\rm t}$)
from the states with opposite (positive) helicity and
parallel alignment of their $z$-component. Therefore their thermal weight is smaller than that of states with negative
helicity and their contribution is not 
sufficient to compensate the positive values of $S_z({\bk}_c)$. Correspondingly, the situation is similar to the case with $k_B T \ll \mu_B(H_z-H_{\rm t})$.

At the topological transition $H_z=H_{\rm t}$ the energy gap closes and quantum critical behavior in the vicinity of this transition is expected~\cite{yang:14,castelnovo:10,gong:12}. In our work we focus exclusively on the mean-field solutions for the superconducting state;  the intricate properties of the transition itself and the concomitant quantum
critical behavior are not the scope of the present work and remain
yet to be explored.

\begin{figure*}[tbp]
\centering
\includegraphics[width=1.99\columnwidth]{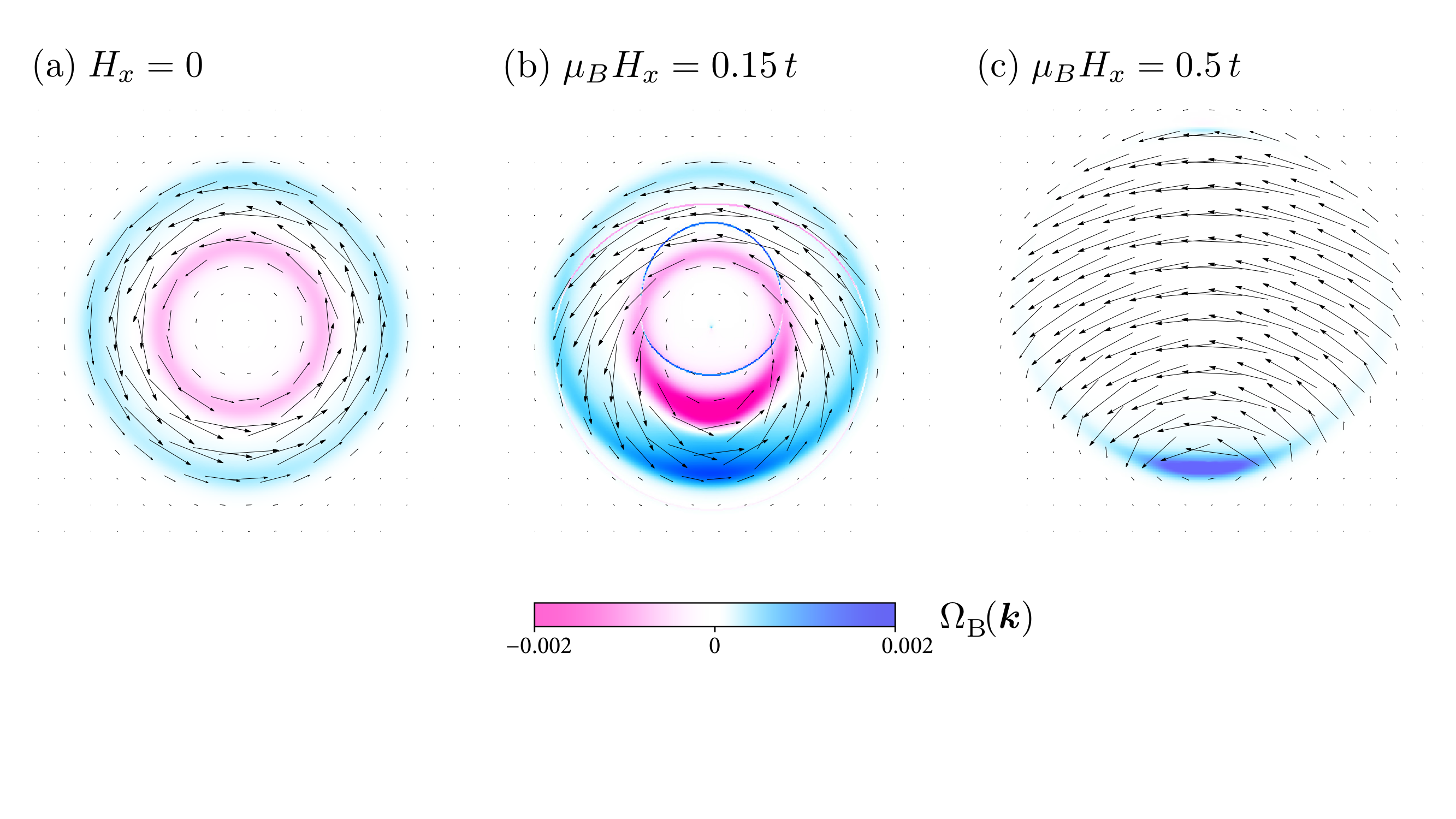}
\vskip-1.5truecm
\caption{(Color online) Vector-fields of the (unnormalized) in-plane spin-expectation values 
$({S}_x(\bk),{S}_y(\bk))$ in the central part of the Brillouin zone 
for a constant out-of-plane field component ($\mu_B H_z=0.1\,t$) 
at zero temperature and
increasing in-plane field strength: 
(a) $H_x=0$, (b) $\mu_B H_x=0.15\,t$, and (c) $\mu_B H_x=0.5\,t$. Here, the critical field strength
 for $\mu_B  H_x$ to reach a topologically non-trivial superconducting state is approximately $0.42\,t$.
The Berry curvature is shown as a colored background. 
In (a) and (b) $N_{\rm S}^{\rm (meron)}\!=0$, while $N_{\rm S}^{\rm (meron)}\!=1$ in (c).
}
\label{Fig7a}
\end{figure*}

\subsection{In-plane magnetic field}
\label{sec:Hxtextures}

As indicated above, finite in-plane components of the magnetic field necessitate an explicit 
solution of the self-consistency condition Eq.~(\ref{selfcondelta}), including the search for 
the optimum COMM. A finite COMM is required, if the in-plane field exceeds a certain 
threshold value~\cite{barzykin02,kaur05,michaeli12,loder15}. This threshold is reached when
the indirect gap closes (see Ref.~\cite{loder15}).
In fact, due to the slight 
difference in magnitude of the offset momenta ${\bq}^+$ and ${\bq}^-$ of the two Fermi 
surface sheets (see Fig.~\ref{FermiSurfaceChirality}) at least two different COMMs are needed 
to optimize the intra-band pairing in the $\xi_{\bk}^+$ and the $\xi_{\bk}^-$ 
band~\cite{michaeli12,loder:13}.

\begin{figure}[t!]
\begin{overpic}
[width=0.8\columnwidth]{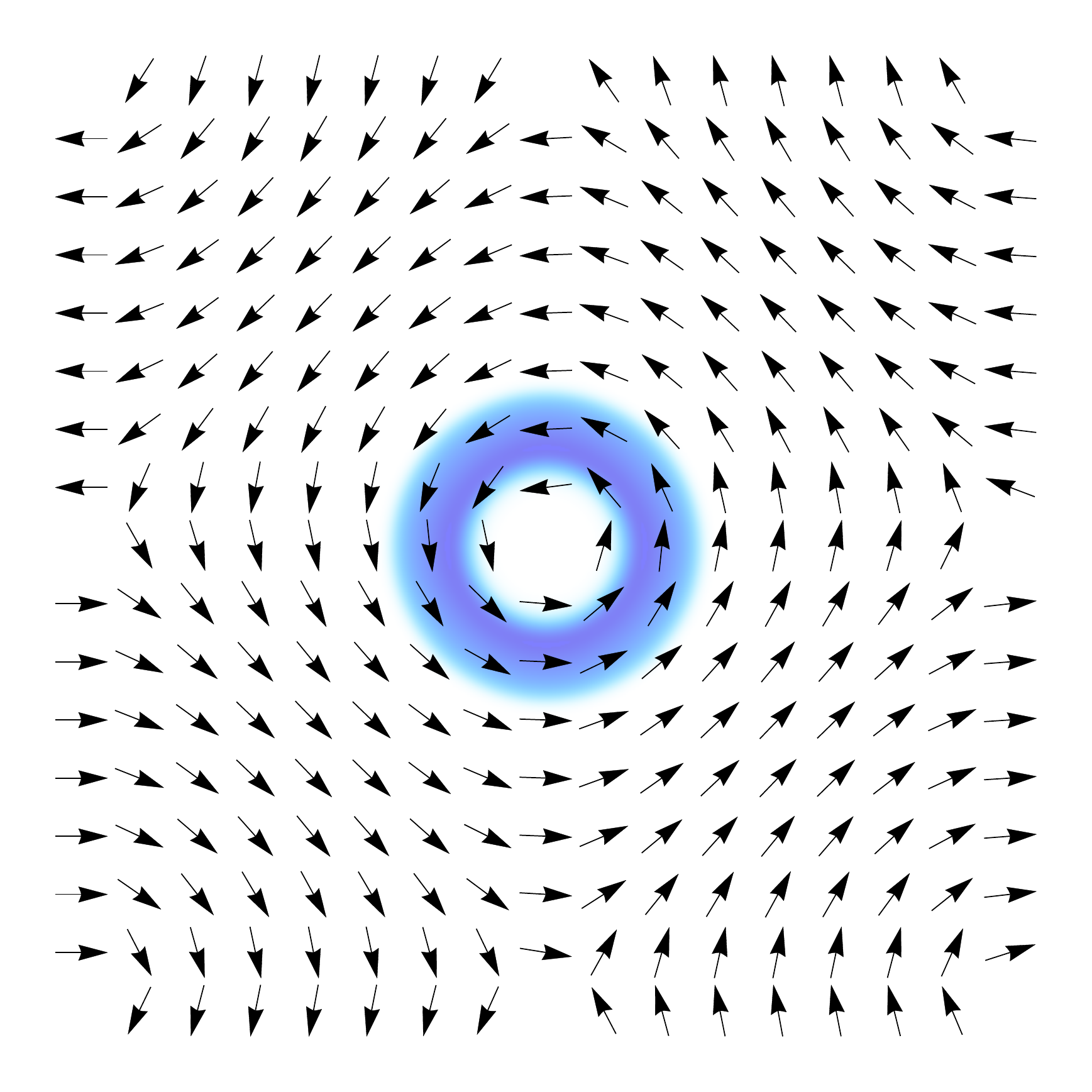}
\put(9,102){(a) $H_x=0$}
\end{overpic}\bigskip
\vspace{2truemm}
\begin{overpic}
[width=0.8\columnwidth]{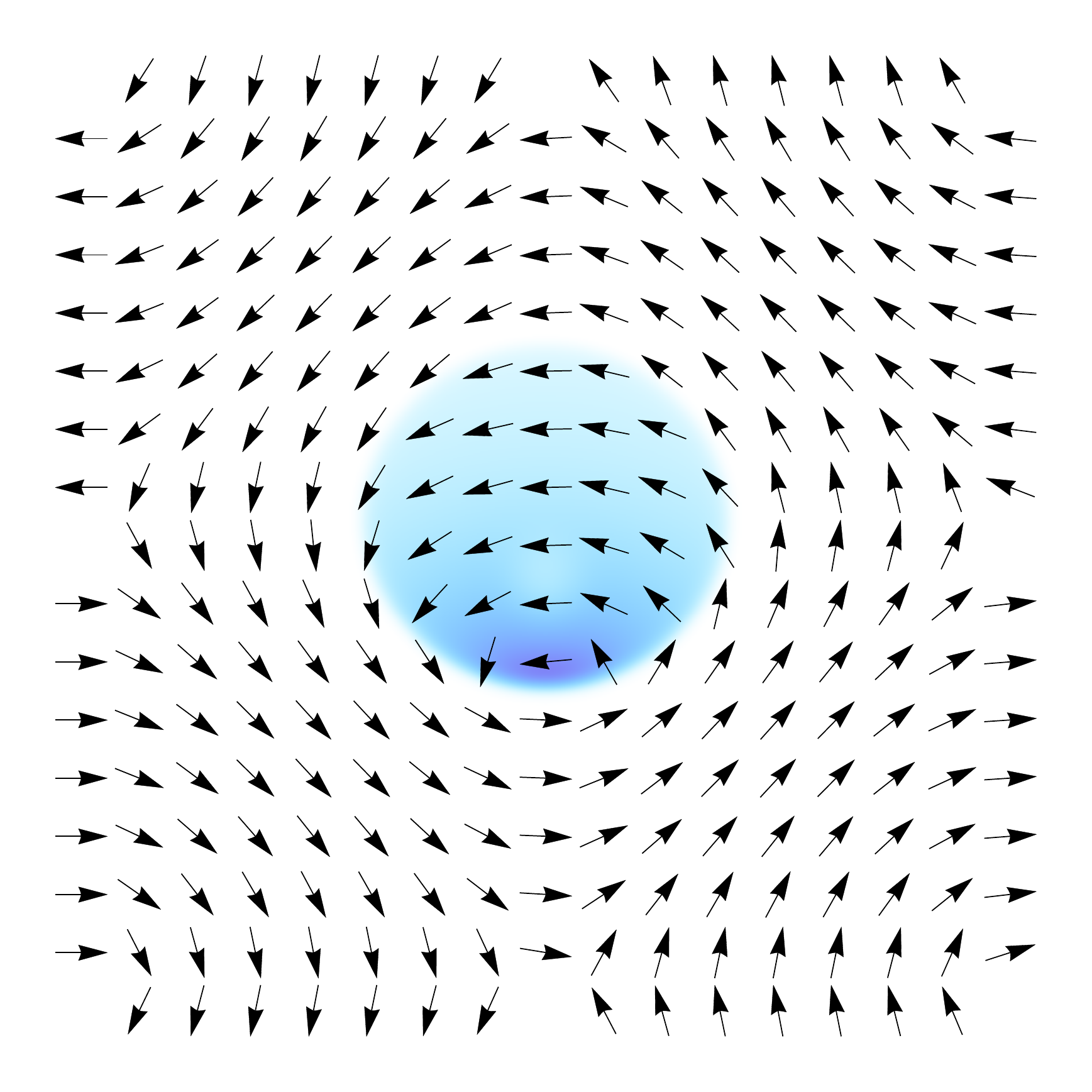}
\put(5,102){(b) $\mu_B H_x=0.5\,t$}
\end{overpic}
\bigskip
\vspace{1truemm}
\begin{overpic}
[width=0.3\columnwidth]{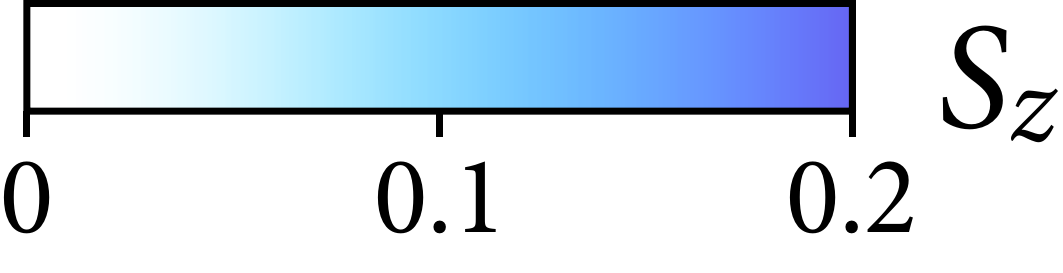}
\end{overpic}
\vspace{3truemm}
\caption{(Color online) Vector fields of the normalized in-plane spin expectation-values, 
$(S_x(\bk),S_y(\bk))/\sqrt{S_x^2+S_y^2}$, in the BZ for (a) the topologically trivial state with 
zero magnetic field, and (b) the topologically non-trivial $s$-wave superconducting state with 
$\mu_B H_x=0.5\,t$ and $\mu_B H_z=0.1\,t$.
The component $S_z(\bk)$ is shown 
as colored background density. The values of $k_x,k_y$ range from $-\pi$ to $\pi$. 
The critical value of $\mu_B H_x$ for the transition to the non-trivial topological state is approximately $\mu_B H_x=0.42\,t$.
}
\label{Fig7}
\end{figure}

For a finite in-plane field component we focus here on the case of zero temperature. A similar reasoning as in Sec.~\ref{sec:outofplaneT} 
also applies to tilted fields for finite temperatures.
As shown in Ref.~\cite{loder15}, the magnetic-field induced topological phase transition 
persists upon rotating the field orientation. Also for an almost in-plane field orientation, the 
momentum-space spin texture characteristically reflects the transition and the meron
counting number $N^{\rm (meron)}_{\rm S}$ discontinuously changes between 0 and $\pm 1$ upon increasing the field strength. The sign of the meron number is determined by the sign of $H_z$. 

Beyond the critical field strength 
where only intraband pairing in the  $\xi_{\bk}^-$ band with a single 
COMM ${\bq}^-$ remains, the relation between $\Omega_{\rm L}({\bk})$
and the Berry curvature $\Omega_{\rm B}({\bk})$, Eq.~(\ref{Omegabar_equal_OmegaBerry}), is modified to
\begin{equation}
\Omega_{\rm B}({\bk})=\frac{1}{2} \left[\Omega_{\rm L} ({\bk}) +
\Omega_{\rm L} (-{\bk}+{\bq}^-)
\right]  \, .
\label{c11}
\end{equation}
This relation signifies that the Berry curvature $\Omega_{\rm B}(\bk)$ 
is a symmetrized version of $\Omega_{\rm L} ({\bk})$ 
and both terms in the bracket of Eq.~(\ref{c11}) integrate up to the same integer $N^{\rm (meron)}_{\rm S}$. Equation~(\ref{c11}) fails for a 
$\k$-dependent order parameter $\Delta_{\bq}(\bk)$ because it will generally break the $\k\leftrightarrow-\k+\q^{-}$ symmetry.

Figure~\ref{Fig7a} shows the evolution of the in-plane spin textures with increasing in-plane field component.
For $H_x=0$ the spin texture is qualitatively equivalent to the texture in Fig.~\ref{Fig3}b for the larger out-of-plane 
field component, but still $H_z< H_{\rm t}$. The colored shaded regions indicate the local Berry curvature as
evaluated from $\Omega_{{\rm B}} ({\bk})$. For $H_x=0$ the COMM of the electron pairs is zero and 
$\Omega_{\rm B}({\bk})=\Omega_{\rm L} ({\bk})$.

For finite in-plane fields, but still below the critical field strength for the topological transition, the Fermi surface sheets of the $\xi_{\bk}^{\pm}$ bands move off-center away from the $\Gamma$-point $\bk ={\bf 0}$ in opposite directions. 
Specifically for the example shown in Fig.~\ref{Fig7a}b, i.e. $H_x>0$, $H_y=0$, the Fermi surface sheets move
along the $k_y$-direction; the pairing on the $\xi_{\bk}^{\pm}$ bands now requires finite COMMs $\bq^+$ and 
$\bq^-$, respectively. The spin texture directly reflects this off-center movement: the region in momentum space around 
which the in-plane spin components wind, shifts along the $k_y$-direction towards the BZ boundary. For $H_x=0$
and  $\mu_B H_x=0.15\,t$ in Figs.~\ref{Fig7a}a and \ref{Fig7a}b, the compensating closed, fuzzy contours with
Berry curvatures of opposite sign remain clearly visible; indeed for these two cases the BZ integral over the Berry
curvature vanishes.

Instead, if $H_x$ exceeds a threshold field to enter the topologically non-trivial superconducting state, the sign
of $\Omega_{{\rm B}}({\bk})$ is unique and its BZ integral leads to the first Chern number $C_{\rm TKNN} =
{\rm sign} (H_z)$. This situation is realized for $\mu_B H_x =0.5\,t$ with the spin texture in Fig.~\ref{Fig7a}c. 
The unique sign of $\Omega_{{\rm B}}({\bk})$ is achieved, when the $\xi_{\bk}^{+}$ band energies are sufficiently 
raised by the tilted Zeeman field such that only the $\xi_{\bk}^{-}$ band states remain occupied. 
Curiously, the winding center for the in-plane spin components has shifted to the near vicinity of the normal-state
Fermi surface of the $\xi_{\bk}^{-}$ band. Of course, in the superconducting phase, a Fermi surface seizes to exist,
but the imprints of the normal state Fermi surfaces are clearly visible in the spin textures. In Figs.~\ref{Fig7a}a and
\ref{Fig7a}b, both the $\xi_{\bk}^{+}$ and the $\xi_{\bk}^{-}$ band are partially occupied, while in Fig.~\ref{Fig7a}c
the $\xi_{\bk}^{+}$ band is empty.

In order to better trace the spin-vector field pattern in momentum space, 
the normalized in-plane spin-expectation values 
$(S_x(\bk),S_y(\bk))/\sqrt{S_x^2+S_y^2}$ are displayed in Fig.~\ref{Fig7}a for the trivial topological phase with out-of-plane
magnetic field $H_z<H_{\rm t}$, and in 
Fig.~\ref{Fig7}b for the topological phase induced by a sufficiently large tilted magnetic field ($\mu_B H_x=0.5\,t$).  
The transition to the topologically non-trivial $s$-wave superconducting state is found at approximately 
$\mu_B H_x=0.42\,t$ for $\mu_B H_z=0.1\,t$. 
The out-of-plane spin component $S_z(\bk)$ is represented
by a colored background.

As already discussed in Sec.~\ref{sec:Hztextures} we identify again 
four micro-vortices in momentum-space also for out-of-plane
fields. It is instructive to count their vorticities  ${\cal V}$ defined through 
the sign of an integral along a loop $\bk(s)$ around the center of the vortex~\cite{footnote1}:
\begin{equation}
{\cal V}(\bk_c) = {\rm sign} \Biggl(
\frac{1}{2\pi}\int_0^{2\pi} \!ds \Bigl[\hat\bS(\bk(s))\times \frac{d\hat\bS(\bk(s))}{ds}  \Bigr]_z 
\Biggr) \, .
\label{vorticity}
\end{equation}
The central vortex in Fig.~\ref{Fig7}a and the micro-vortex at the BZ corners
${\bk_c = (\pm\pi, \pm\pi)}$ and ${\bk_c =\bf (\mp\pi, \pm\pi)}$ have vorticity $+1$ whereas the two micro-vortices at the BZ faces
${\bk_c = (0, \pm\pi)}$ and ${\bk_c = (\pm\pi, 0)}$ have each vorticity $-1$. The total vorticity $\cal V_{\rm t}$ is zero. In fact, for periodic boundary conditions, it is a necessary requirement to have ${\cal V_{\rm t}} =0$, i.e., for a BZ represented by a torus. This implies, for example, that the central vortex cannot appear alone. Indeed, from the fact that the torus is mapped by $\cal M$ to a unit sphere it is evident that four micro-vortices must exist which are mapped to the south pole of the sphere (see Fig.~\ref{Torus_Sphere}b).

For the topologically non-trivial $s$-wave superconducting state in a tilted magnetic field (see Fig.~\ref{Fig7}b), the micro-vortices at the BZ boundary are preserved. The central vortex, however, is moved towards the normal state Fermi surface in the direction perpendicular to the in-plane magnetic field (as discussed in connection with Fig.~\ref{Fig7a}). But the center of the central vortex never traverses the normal-state Fermi surface. The vortex center has the maximum $S_z$ and is therefore mapped by $\cal M$ to the north pole of the sphere for magnetic fields larger than the critical field, but the vorticity of this spin pattern is preserved and the total vorticity remains zero. The important observation here is that the central vortex is trapped inside the Fermi surface when the superconducting gap is finite. In the normal state at small temperature 
the central vortex glides outside the Fermi surface and moves to 
the BZ boundary for $H_x>\alpha|\sin k_{y,\t F}|$, where $k_{y,\t F}$ is the Fermi momentum in the direction of $(0,-\pi)$. The spin field vortex pattern is thereby annihilated. Whether a similar trapping of the central vortex is possible also in a nodal superconductor may be yet another issue for future work on momentum-space spin textures.   

\section{Conclusion}
\label{sec:conclusion}

As we have demonstrated for the specific example of a topological $s$-wave superconductor, the spin texture in momentum space provides an alternative tool to identify a system's topological character. The meron counting number is a well-defined topological
invariant; its discontinuous change signifies a transition between topologically distinct ground states. For the meron number to be a 
topological invariant, it proves crucial that the spin texture has isolated singular points at the BZ boundary. These singularities
control the essential maps from the compact manifold of the BZ to the unit sphere of normalized spin-expectation values and
ultimately allow for a finite, integer valued meron number.

For time-reversal invariant topological insulators it was shown in Sec.~\ref{sec:concepts} that the skyrmion counting number of the spin texture
is an integer valued topological invariant, equivalent to the Brouwer degree of mapping  or the first Chern number, 
which is the BZ integral over the Berry curvature. The
topological $s$-wave superconductor relies on the Zeeman coupling to an external magnetic field. This time-reversal symmetry
breaking magnetic field prevents the spin map to cover the full spin unit sphere and therefore necessitates to turn to integer valued meron or half-integer valued skyrmion numbers. The natural question still has to be answered, whether this paradigm applies to 
topological insulators with broken time-reversal symmetry, too.

An example, in this context, is the emergence of a hedgehog spin-texture in momentum space in magnetically doped 3D topological
insulators~\cite{xu:12}. 
Also on the surface of a 3D topological insulator, if an exchange field is applied, momentum-space hedgehog or 
skyrmion textures emerge~\cite{mohanta:16}, where the former has skyrmion number $1/2$.

In contrast to the ground state concepts of Berry curvature or Chern number, the topological fingerprint in a system's
spin texture may allow a natural extension to finite temperatures. As we have shown for the topological $s$-wave superconductor,
the spin expectation values at finite temperatures already reveal signatures of the groundstate with finite integer valued
topological invariants. This raises the general question, to what extent topological phase transitions influence the electronic properties
at finite temperatures near the critical control parameters for the topological transition. The spin textures, which we have 
encountered and analyzed in this work, appear to offer an example for such a phenomenon, and remains yet to be 
translated into a broader context.

\section*{Acknowledgements} 
We gratefully acknowledge  Alexander Herrnberger's support in the design of figure 6. This work was supported by the DFG through TRR 80.

\bigskip 

\appendix

\section{Brouwer degree and skyrmion number}{\label{app1}}

In this section we derive Eqs.~\eqref{brouwer1} and \eqref{Om1} of the main text. The domain of the map $\bh(\bk)$ is $T^2$ and a subset of $\Rr^3$ is its range. This subset is homeomorphic to a subset of $S^2$, the range of the normalized map $\bh/h$. Both $\bh$ and $\bh/h$ have the same Brouwer degree. 

We select one of the points $\bk_l$, say $\bk_0$, and---for convenience---shift coordinates such that $\bk_0=\bnull$. 
Then $h(\bnull)=h_z(\bnull)$, and $h_x(\bnull)=h_y(\bnull)=0$. The two components of the map, $h_{x}(k_1,k_2)$ and $h_{y}(k_1,k_2)$, are 
linearized around $\bk=\bnull$ as
  \beq
  h_x=J_{11}k_1+J_{12}k_2,\quad h_y=J_{21}k_1+J_{22}k_2,
  \eeq
where $J_{ik}$ denotes the elements of the Jacobi matrix
  \beq
  \hJ(\bk_0)=\frac{\pa(h_x,h_y)}{\pa(k_1,k_2)}\biggl |_{\bk_0}
  \eeq
at the point $\bk_0$, i.e.\ the differential of the map from $T^2$ into $\Rr^3$ given by 
$\bh(\bk)$, where we have chosen $h_x(\bk)$, $h_y(\bk)$ as local coordinates of the image 
manifold. We thereby find
  \beq
  R^2(k_1,k_2)=h_x^2+h_y^2 + {\cal O}(k_j^4),
  \eeq
  \beq
  v_1=\frac{J(\bk_0)}{h_x^2+h_y^2}k_2, \quad v_2=-\frac{J(\bk_0)}{h_x^2+h_y^2}k_1,
  \eeq
with the Jacobian $J(\bk_0)=\det\hJ(\bk_0)$. To evaluate $\oint_{c_l}\bv$ for a small circle 
$c_0$ with radius $r$ around $\bk_0=\bnull$, we set $k_1=r\cos\psi$, $k_2=r\sin\psi$. With
$\rd\bs=(-k_2\rd\psi,k_1\rd\psi)^\top$, we obtain
  \beq
  \oint_{c_0}\bv=
  -J(\bk_0)r^2\int_0^{2\pi}\frac{\rd\psi}{h_x^2(r,\psi)+h_y^2(r,\psi)}.
  \label{vint}
  \eeq
We write the denominator of Eq.~(\ref{vint}) as
  \beq
  h_x^2+h_y^2=\bk^\top\hJ^\top(\bk_0)\hJ(\bk_0)\bk,
  \eeq
where the quadratic form $\hJ^\top\hJ$ can be diagonalized by an orthogonal transformation. 
With the eigenvalues $\lam_{1,2}$ of $\hJ(\bk_0)$ and upon rotating $\psi$ accordingly, we are lead to
  \beq
  h_x^2+h_y^2=r^2(\lam_1^2\cos^2\psi+\lam_2^2\sin^2\psi).
  \eeq
Performing the elementary integral over the angle $\psi$
we find
  \beq
  \oint_{c_0}\bv=-2\pi\frac{J(\bk_0)}{|\lam_1\lam_2|}=- 2\pi\ \sign(J(\bk_0)). 
  \eeq
    Upon collecting all preimages $\bk_l$ of $(0,0,h_z)$ for arbitrary $h_z$, i.e. all singular points 
of $\bv$, the final result reads
  \beq
  \int_{BZ}\bOm=2\pi\sum_l\sign(J(\bk_l))=2\pi C_{\rm Brouwer}.
  \label{brouwer-ap}
  \eeq
To derive Eq.~\eqref{Om1}, we evaluate \eqref{BerryOm} using \eqref{conn}
and obtain
  \begin{widetext}
    \beq
  \Om=2R^{-2}[(\pa_1h_y)(\pa_2h_x)-(\pa_1h_x)(\pa_2h_y)]+
  (h_y\pa_2h_x-h_x\pa_2h_y)\pa_1\left(R^{-2}\right)
  - (h_y\pa_1h_x-h_x\pa_1h_y)\pa_2\left(R^{-2}\right).
  \label{curv}\eeq
  \end{widetext}
The terms in Eq.~\eqref{curv}, which contain derivatives of $R^{-2}$, can be combined into
\beq
  \frac{2}{hR^4}\left\{(h_x^2+h_y^2)(2h-h_z)J_{xy}+(h-h_z)^2[h_xJ_{yz}+h_yJ_{zx}]\right\}
  \label{partr}
\eeq
where we have used the notation Eq.~\eqref{nota} in the main text.
The expression in \eqref{partr} is still divergent at the points $\bk_l$ where $R$ vanishes. 
However, together with the first term in Eq.~\eqref{curv} we find
\begin{align}
  \Om = & J_{xy}\left[\frac{2}{hR^4}(2h-h_z)(h_x^2+h_y^2)-\frac{2}{R^2}\right]\nn\\
  &+\frac{2}{hR^4}(h-h_z)^2[h_xJ_{yz}+h_yJ_{zx}],
\end{align}
which yields after some simplifications Eq.~\eqref{Om1}.

To show that $N_S=C_{\rm Brouwer}$,
we first write $\bh(\bk)$ in spherical coordinates: $\bh(\bk)=h(\bk)\be_r(\vth({\bk}),\vp({\bk}))$, and
\beq
\pa_j\bh=(\pa_jh)\be_r+h(\pa_j\vth)\be_\vth+h\sin\vth(\pa_j\vp)\be_\vp.
\eeq
The cross product in the skyrmion density Eq.~(\ref{skyrmdensity}) then takes the form
\begin{align}
  \pa_1\bh&\times\pa_2\bh = h^2\sin\vth\,(\pa_1\vth\, \pa_2\vp  - \pa_1\vp\,\pa_2\vth)\,\be_r\nn\\
  &+h\sin\vth\,(\pa_1\vp\,\pa_2h  -\pa_1h\,\pa_2\vp)\,\be_\vth\\
  &+h\,(\pa_1h \,\pa_2\vth\, -\pa_1\vth\,\pa_2h)\,\be_\vp,\nn
\end{align}
and therefore
\beq
\Om_{\rm h}(\bk)=\sin\vth[(\pa_1\vth)(\pa_2\vp)-(\pa_1\vp)(\pa_2\vth)].
\label{jacobian}
\eeq
Note that the derivatives of $h(\bk)$ have dropped out in Eq.~(\ref{jacobian}). This implies
\beq\label{omega_norm}
\Om_{\rm h}(\bk)=\frac{\bh}{h^3}\cdot[\pa_1\bh\times\pa_2\bh]
=\hat{\bh}\cdot[\pa_1\hat{\bh}\times\pa_2\hat{\bh}],
\eeq
with the normalized $\hat{\bh}=\bh/h$. 
The right hand side of Eq.~(\ref{jacobian}) is $\sin\vth({\bk})$ times the Jacobian
\beq
J(\bk)=\frac{\pa(\vth,\vp)}{\pa(k_1,k_2)}
\eeq
of the map $\hat{\bh}(\bk)$ from the torus to the unit sphere. For $N_{\rm S}$ of 
Eq.~(\ref{NS}) we eventually find
\beq
4\pi N_{\rm S}=\!\!\int_{BZ}\!\!\sin\vth({\bk})\, J(\bk)\,\rd^2 k\!=\!N_{\rm B}\!\int_{S^2}\!\!\sin\vth\,\rd\vth\rd\vp=4\pi N_{\rm B}
\label{brouwer2}
\eeq
where $N_{\rm B}\in\Zz$ counts the number of times each (regular) point of $S^2$ is attained by the map 
$\hat{\bh}(\bk)$; the sign of $N_{\rm B}$ reflects the orientation of the map. 
$N_{\rm S}$ is therefore again the Brouwer degree of $\hat{\bh}(\bk)$
and Eq.~\eqref{brouwer2} is equivalent to Eq.~\eqref{brouwer1} because the surface form 
$\sin\vth\,\rd\vth\wedge\rd\vp$ has the same orientation as $\rd h_x\wedge\rd h_y$ in the vicinity 
of the singular points with $\hat{\bh}=\be_z$.

Yet another possibility to compute the skyrmion number $N_{\rm S}$ is furnished by the quantity
\beq
\bar{\bS_\lam}=\left(\frac{h_x}{\bar{h}},\frac{h_y}{\bar{h}},\lam\frac{h_z}{h}\right)
\eeq
for arbitrary real $\lam$ with $\bar{h}=\sqrt{h_x^2+h_y^2}$. To verify this option, we begin with $\lam=1$ 
and note that $\bar{\bS}_1$ is the horizontal projection of the point 
$\hat{\bh}=(\cos\vp\sin\vth,\sin\vp\sin\vth,\cos\vth)$ on the unit sphere, onto the point
$(\cos\vp,\sin\vp,\cos\vth)$ on the cylinder with unit barrel radius. Since the surface elements
$\sin\vth\rd\vth\wedge\rd\vp$ of the sphere and $\rd\vp\wedge\rd z=\rd\vp\wedge\rd(\cos\vth)$ of the cylinder are 
equivalent, we have 
\beq\label{Sbar_product}
\bar{\bS}_1\cdot[\pa_1\bar{\bS}_1\times\pa_2\bar{\bS}_1]=
\frac{\pa(\vp,z)}{\pa(k_1,k_2)}=\sin\vth J(\bk).
\eeq
where 
\beq\label{Sbar_1}
\bar{\bS_1}=\left(\frac{h_x}{\bar{h}},\frac{h_y}{\bar{h}},\frac{h_z}{h}\right)
=\left(\frac{S_x}{\bar{S}},\frac{S_y}{\bar{S}},\frac{S_z}{S}\right)
\eeq
with $\bar{S}=\sqrt{S_x^2+S_y^2}$ and $S=\sqrt{S_x^2+S_y^2 +S_z^2}$. Here 
the spin expectation values have been introduced as in Eq.~(\ref{spintexture}).

The equality~(\ref{Sbar_product}) implies
\beq
N_{\rm S}=\frac{1}{4\pi}\int_{BZ}\bar{\bS}_1\cdot[\pa_1\bar{\bS}_1\times\pa_2\bar{\bS}_1]\,\rd^2 k \,.
\eeq
As the surface element of the cylinder scales with $\lam$, we immediately generalize to arbitrary $\lam$
\beq
\bar{\bS}_\lam\cdot[\pa_1\bar{\bS}_\lam\times\pa_2\bar{\bS}_\lam]=\lam\sin\vth J(\bk).
\eeq
With the Berry curvature, written in this geometry, 
\beq
\Omega_{\lambda}(\bk) =\bar{\bS}_\lam\cdot[\pa_1\bar{\bS}_\lam\times\pa_2\bar{\bS}_\lam],
\eeq
the skyrmion number is reexpressed as
\beq
N_{\rm S}=\frac{1}{4\pi\lam}\int_{BZ}\bar{\bS}_\lam\cdot[\pa_1\bar{\bS}_\lam\times\pa_2\bar{\bS}_\lam]
\,\rd k_1\rd k_2.
\eeq

\section{Pairing amplitudes in the helicity basis}
\label{app0}
With the shorthand notation $h({\bk})^{-}=h_x({\bk})-ih_y({\bk})$ and $h_z=h_z({\bk})$ the
intra- and inter-band pairing amplitudes in the helicity basis are explicitly given by
\begin{widetext}
\begin{eqnarray}
\Delta^{++}({\bk},{\bq}) &=& \displaystyle{\Delta_{\bq}\over 2\sqrt{h({\bk})h(-{\bk}+{\bq})}}
\left\{ \sqrt{\displaystyle{h({\bk})+h_z\over h(-{\bk}+{\bq})+h_z}} h^-(-{\bk}+{\bq}) -
\sqrt{\displaystyle{h(-{\bk}+{\bq})+h_z\over h({\bk})+h_z}} h^-(\bk)
\right\} \, ,\\
\Delta^{--}({\bk},{\bq}) &=& \displaystyle{\Delta_{\bq}\over 2\sqrt{h({\bk})h(-{\bk}+{\bq})}}
\left\{ \sqrt{\displaystyle{h(-{\bk}+{\bq})-h_z\over h({\bk})-h_z}} h^-(\bk) -
\sqrt{\displaystyle{h({\bk})-h_z\over h(-{\bk}+{\bq})-h_z}} h^-(-{\bk}+{\bq})
\right\} \, ,\\
\Delta^{+-}({\bk},{\bq}) &=& \displaystyle{\Delta_{\bq}\over 2\sqrt{h({\bk})h(-{\bk}+{\bq})}}
\left\{ \sqrt{\displaystyle{h({\bk})+h_z\over h(-{\bk}+{\bq})-h_z}} h^-(-{\bk}+{\bq}) +
\sqrt{\displaystyle{h(-{\bk}+{\bq})-h_z\over h({\bk})+h_z}} h^-(\bk)
\right\} \, ,\\
\Delta^{-+}({\bk},{\bq}) &=& \displaystyle{-\Delta_{\bq}\over 2\sqrt{h({\bk})h(-{\bk}+{\bq})}}
\left\{ \sqrt{\displaystyle{h(-{\bk}+{\bq})+h_z\over h({\bk})-h_z}} h^-(\bk) +
\sqrt{\displaystyle{h({\bk})-h_z\over h(-{\bk}+{\bq})+h_z}} h^-(-{\bk}+{\bq})
\right\} \, .
\end{eqnarray} 
\end{widetext}
Note that in general $\Delta^{+-}({\bk},{\bq}) \neq \Delta^{-+}({\bk},{\bq})$. Only for the 
special $H_z =h_z =0$, i.e. for an in-plane magnetic field, $\Delta^{+-}({\bk},{\bq})=
\Delta^{-+}({\bk},{\bq})$ and $\Delta^{+-}({\bk},{\bq})$ is even with respect to 
interchanging ${\bk}$ and $-{\bk}+{\bq}$.


\begin{references}
\bibitem{footnote2} Note the missing factor $\sqrt{-1}$ in the definition of the Chern class as compared to~\cite{jost}, p.127. This is due to our definition of the connection $\bv$ as real-valued in Eq.\eqref{connection}.
\bibitem{footnote1} We emphasize that ${\cal V}$ is not the winding number but rather distinguishes vortex 
and anti-vortex. The definition for the vorticity is valid for an $xy$-model even without the enclosing sign-function.
Here we need the sign-function for a loop in the vicinity of the central vortex in the topologically non-trivial state 
where the spins have a sizable $S_z$ expectation value.
\end{references}
\end{document}